\documentstyle[12pt,amsmath,amssymb,epic]{article}

\newcommand{\bbbone}{\mathchoice {\rm 1\mskip-4mu l} {\rm 1\mskip-4mu l}
{\rm 1\mskip-4.5mu l} {\rm 1\mskip-5mu l}}
\newcommand{\scalprod}[2]{\left\langle {#1}, {#2}\right\rangle}
\newcommand{\dom}{{\cal D}}

\newcommand{\IM}{{\rm Im}}
\newcommand{\fer}[1]{(\ref{#1})}
\newcommand{\ran}{{\rm Ran\,}}
\renewcommand{\ker}{{\rm Ker\,}}
\newcommand{\repsilonbar}{\,\overline{\!R}_\epsilon}
\newcommand{\repsilon}{R_\epsilon}

\newcommand{\edeltanot}{E_\Delta^0}
\newcommand{\chideltanot}{\chi_\Delta^0}
\newcommand{\h}{{\cal H}}
\newcommand{\cx}{{\mathbb C}}
\newcommand{\r}{{\mathbb R}}

\renewcommand{\d}{{\rm d}}
\newcommand{\Pibar}{\overline{\Pi}}
\newcommand{\Pbar}{\overline{P}}

\newcommand{\av}[1]{\left\langle{#1}\right\rangle}

\newcommand{\mm}{{\frak M}}

\newcommand{\cc}{{\cal C}}

\newcommand{\tr}{{\rm tr\,}}

\newcommand{\Obl}{{\Omega_{\beta,\lambda}}}
\newcommand{\Obz}{{\Omega_{\beta,0}}}
\newcommand{\obl}{{\omega_{\beta,\lambda}}}
\newcommand{\obz}{{\omega_{\beta,0}}}
\newcommand{\oblL}{{\omega^\Lambda_{\beta,\lambda}}}
\newcommand{\obzL}{{\omega^\Lambda_{\beta,0}}}
\newcommand{\HlL}{H^\Lambda_\lambda}

\newcommand{\HfL}{H_f^\Lambda}
\newcommand{\hfL}{h_f^\Lambda}
\newcommand{\HzL}{H_0^\Lambda}
\newcommand{\HoL}{H_0^\Lambda}

\newcounter{resultcounter} 
\stepcounter{resultcounter}

\newtheorem{theorem}{Theorem}[section]

\newtheorem{proposition}{Proposition}[section]

\begin{document}

\setcounter{page}{0}

\title{\vspace*{-1cm}
Another return of \\
``Return to Equilibrium''}

\author{
J\"urg Fr\"{o}hlich\footnote{juerg@itp.phys.ethz.ch}\ \ \  and \  
Marco Merkli\footnote{present address: Department of Mathematics and Statistics, McGill University, 805 Sherbrooke W., Montreal, Quebec, Canada, H3A 2K6, merkli@math.mcgill.ca}\ \\      
Theoretical Physics\\
 ETH- H\"{o}nggerberg \\ 
CH-8093 Z\"{u}rich, Switzerland\\
}
\date{\today}
\maketitle

\begin{abstract}
The property of ``{\it return to equilibrium}'' is established for a class of quantum-mechanical models describing interactions of a (toy) atom with black-body radiation, or of a spin with a heat bath of scalar bosons, under the assumption that the interaction strength is {\it sufficiently weak}. For models describing the first class of systems, our upper bound on the interaction strength is {\it independent} of the temperature $T$, (with $0<T\leq T_0<\infty$), while, for the spin-boson model, it tends to zero logarithmically, as $T\rightarrow 0$. Our result holds for interaction form factors with physically realistic infrared behaviour.\\
\indent
Three key ingredients of our analysis are: a suitable concrete form of the Araki-Woods representation of the radiation field, Mourre's positive commutator method combined with a recent virial theorem, and a norm bound on the difference between the equilibrium states of the interacting and the non-interacting system (which, for the system of an atom coupled to black-body radiation, is valid for {\it all} temperatures $T\geq 0$, assuming only that the interaction strength is sufficiently weak). 
\end{abstract}
\thispagestyle{empty}
\setcounter{page}{1}
\setcounter{section}{1}


\setcounter{section}{0}

\section{Introduction}
The problem of {\it return to equilibrium} for models describing small systems with finitely many degrees of freedom coupled to a dispersive heat bath at positive temperature has been studied at various levels of mathematical precision, since the early days of quantum theory. Fairly recently, a new approach to this problem based on spectral theory for thermal Hamiltonians, or Liouvillians, has been described and applied to simple models in [JP1, JP2]. The general {\it strategy} followed in our paper is based on the spectral approach proposed in these references; but our {\it tactics} are quite different and draw inspiration from techniques developed in [M] that have been motivated by methods in [BFSS]. For further results and methods relevant to our paper, see [BFS, DJ2, GGS1] and, in particular, [FM, FMS]. The work described in {\it all} these references relies on the deep insights of Haag, Hugenholtz and Winnink [HHW] and on the Araki-Woods representation [AW].\\
\indent
The main result proven in this paper is Theorem \ref{rtethm}, asserting {\it return to equilibrium} for a class of models describing a ``small system'' with a finite-dimensional state space coupled to a ``large system'', a dispersive heat bath at some temperature $T$, with $0<T\leq T_0<\infty$. The heat bath is modelled by a spatially infinitely extended free massless bosonic field. The systems we consider fall into two categories corresponding to a {\it regular} or a {\it singular} infrared behaviour of the coupling between the two subsystems. Both cases are {\it physically realistic}.\\
\indent
We show return to equilibrium under the assumption that the interaction strength is sufficiently weak. For infrared-regular systems, such as toy atoms interacting with black-body radiation, our upper bound on the interaction strength only depends on $T_0$, but not on $T<T_0$. For infrared singular systems, such as the usual spin-boson model, our upper bound on the interaction strength depends on $T$ and tends to zero logarithmically, as $T\rightarrow 0$. \\
\indent
The proof of Theorem \ref{rtethm}, which is presented in Section \ref{rtethmproof}, relies on a result of independent interest, Theorem \ref{betauniform}, which says that the norm of the difference of the equilibrium states of the coupled system and of the non-interacting system is small, for weak interaction strength (uniformly in the temperature in the infrared-regular case, and with an explicit temperature-depent upper bound on the interaction strength for the singular case). Theorem \ref{betauniform} is proven in Section \ref{betauniformsect}, and the proof draws on ideas developed in [A, FNV, BFS, F].\\
\indent
With Theorems \ref{rtethm} and \ref{betauniform}, we accomplish two goals. First, for infrared-regular systems, our results are {\it uniform} in the temperature $T$, for $0<T\leq T_0$ (where the high-temperature bound, $T_0$, has a clear physical origin, see also the comment after Theorem \ref{rtethm} below), assuming only that the interaction strength is small enough, with an upper bound only depending on $T_0$. Second, our results also hold for infrared-singular systems, provided the temperature is not too small (depending on the interaction strength). \\
\indent
In order to render our discussion more concrete, we describe the models studied in this paper more explicitly. The first class describes systems consisting of an atom, or of an array of finitely many atoms, coupled to the quantized electromagnetic field. We assume that the temperature $T$ of the electromagnetic field is so small that it is justified to treat the atomic nucleus as {\it static} and to neglect the role played by atomic states of high energy, in particular those corresponding to the continuous energy spectrum describing an ionized atom. Thus, the upper bound, $T_0$, on the temperature range considered in this paper is determined by the requirement that
\begin{equation}
k_BT_0 <\!\!<  m_{at}c^2, \ \ \ k_BT_0<\Sigma,
\label{j1}
\end{equation}
where $k_B$ is Boltzmann's constant, $m_{at}c^2$ is the rest energy of an atom, and $\Sigma$ is the ionization(-threshold) energy. If $T\leq T_0$, with $T_0$ satisfying \fer{j1}, then an atom can be described, approximately (in the spirit of the Born-Oppenheimer approximation), as a quantum-mechanical system with a {\it finite-dimensional state space} spanned by those unperturbed atomic eigenstates corresponding to atomic energies $E\lesssim const\ k_BT_0$ in the discrete spectrum. This defines what we call a ``toy (or truncated) 
atom''. \\
\indent
The coupling of the toy atom to the quantized radiation field is described, in the dipole approximation, by an interaction term
\begin{equation}
-e  \vec d_{at}\cdot\vec E(\rho),
\label{j2}
\end{equation}
where $e$ is the elementary electric charge, $\vec d_{at}$ is the atomic dipole (moment) operator, and $\vec E$ is the quantized electric field. Furthermore, $\rho$ is a density function corresponding to an approximate (smooth) $\delta$-function peaked at the position of the nucleus and of width comparable to the size of the atom. (The interaction term \fer{j2} defines the Ritz Hamiltonian.) When expressed in terms of (Newton-Wigner) photon creation- and annihilation operators the interaction term \fer{j2} gives rise to a momentum-space form factor $g_0(k)$ (see Section \ref{modelsection}) corresponding to
\begin{equation}
g_0(k)=i \sqrt{|k|}\widehat\rho (k) \propto \sqrt{|k|}, 
\label{j3}
\end{equation}
for $|k|\rightarrow 0$, where $k$ is the photon momentum. Interactions characterized by an infrared behaviour $g_0(k)\propto |k|^p$, as $|k|\rightarrow 0$, with $p>-1/2$, are called {\it infrared-regular}. Nowhere in our analysis will the {\it helicity of photons} play an interesting role. The helicity- (polarization-) index will therefore be suppressed in our notation, and we shall think of the heat bath as being described by a scalar field (instead of a transverse vector field). \\
\indent
The second class of models deals with systems of a quantum mechanical spin $\vec S$, with $\vec S\cdot \vec S=s(s+1)$ (and usually $s=1/2$) coupled to a heat bath described in terms of a quantized, real, massless scalar field $\varphi$. Before the spin is coupled to the heat bath it exhibits precession around an external field $\vec B$ pointing in the $z$-direction. Its dynamics is generated by a Hamiltonian
\begin{equation}
H_{spin}=\epsilon S_z,\ \ \ \mbox{with \ \ \ $\epsilon\propto |\vec B|$}.
\label{j4}
\end{equation}
The interactions of the impurity spin with the heat bath give rise to spin-flip processes described by an interaction term e.g. of the form
\begin{equation}
g S_x\varphi(\rho),
\label{j5}
\end{equation}
where $g$ is a coupling constant, and $\rho$ is a density function as described above. The bound, $T_0$, on the temperature range considered is determined by our desire not to take orbital excitations of the particle (an electron, neutron or atom in a dispersive medium, such as an insulator) carrying the impurity spin $\vec S$ into account.\\
\indent
When $\varphi$ is expressed in terms of (Newton-Wigner) creation- and annihilation operators the interaction term \fer{j5} gives rise to a momentum-space form factor $g_0$, with 
\begin{equation}
g_0(k)=\frac{\widehat\rho(k)}{\sqrt{|k|}}\propto\frac{1}{\sqrt{|k|}},
\label{j6}
\end{equation}
for $|k|\rightarrow 0$, where $k$ is the momentum of a scalar boson in the heat bath. Interactions characterized by an infrared behaviour \fer{j6} are called {\it infrared-singular}.\\
\indent
The physical interest of the second model, the {\it spin-boson model}, is somewhat limited. But it has often been used to illustrate the phenomena of interest to us in this paper. \\

A general class of model systems reminiscent of the ones just described is introduced, in a formal mathematical way, in Section \ref{modelsection} below. In the following, we attempt to clarify what we mean by ``return to equilibrium''. Let $\cx^d$ be the state space of the ``small system'' (the toy atom or impurity spin), and let ${\cal B}(\cx^d)$ denote the algebra of matrices acting on $\cx^d$. Let $\frak W$ denote the algebra of Weyl operators over a suitably chosen space of one-boson test functions describing the quantum-mechanical degrees of freedom of the heat bath. The Weyl operators, which are exponentials of field operators smeared out with test functions, are bounded operators, and the algebra $\frak W$ they generate is a $C^*$-algebra. The kinematics of the composed system consisting of the ``small system'' and the heat bath is described by the $C^*$-algebra
\begin{equation}
{\frak A}={\cal B}(\cx^d)\otimes{\frak W},
\label{j7}
\end{equation}
and its dynamics, in the Heisenberg picture, is given by a one-parameter group $\{\alpha_t\}$, with $t\in\r$ denoting time, of $*$automorphisms of $\frak A$. Before the small system is coupled to the heat bath, $\alpha_t\equiv \alpha_{t,0}$ is given by 
\begin{equation}
\alpha_{t,0}=\alpha_t^{at}\otimes \alpha_t^f
\label{j8}
\end{equation}
where $\alpha_t^{at}(A)=e^{itH_{at}} A e^{-itH_{at}}$, $A\in{\cal B}(\cx^d)$, is the Heisenberg-picture dynamics of an isolated toy atom, $H_{at}$ is its Hamiltonian, and where $\alpha_t^f$ describes the Heisenberg-picture dynamics of the heat bath. We choose $\{\alpha_t^f\}$ to be the $*$automorphism group of $\frak W$ describing the dynamics of free, relativistic, massless bosons, such as photons (but, as announced, we shall suppress reference to their helicity in our notation).\\
\indent
Let $\omega_\beta^{at}$ and $\omega_\beta^f$ be the equilibrium states of the small system isolated from the heat bath, and of the free heat bath, respectively, at inverse temperature $\beta=(k_B T)^{-1}$. Let $\h$ denote the Hilbert space of state vectors of the composed system obtained from the algebra $\frak A$ in \fer{j7} and the equilibrium state, $\omega_{\beta,0}$, given by 
\begin{equation}
\omega_{\beta,0}=\omega_\beta^{at}\otimes \omega_\beta^f,
\label{j9}
\end{equation}
before the small system is coupled to the heat bath, by applying the GNS construction. Furthermore let $\Obz\in\h$ denote the cyclic vector in $\h$ corresponding to the state $\obz$, and let $\pi_\beta$ be the GNS representation of $\frak A$ on $\h$. Since $\obz$ is time-translation invariant, in the sense that $\obz(\alpha_{t,0}(A))=\obz(A)$, for all $A\in\frak A$ and all times $t\in\r$, there is a selfadjoint operator, $L_0$, called thermal Hamiltonian or {\it Liouvillian}, acting on $\h$ with the properties
\begin{equation}
\pi_\beta(\alpha_{t,0}(A)) =e^{itL_0} \pi_\beta(A) e^{-itL_0},
\label{j10}
\end{equation}
for all $A\in\frak A$, and 
\begin{equation}
L_0\Obz=0.
\label{j11}
\end{equation}
In order to describe interactions between the small system and the heat bath at inverse temperature $\beta$, one replaces the (unperturbed) Liouvillian $L_0$ by an (interacting) Liouvillian $L_\lambda$, which is a selfadjoint operator on $\h$ given by 
\begin{equation}
L_\lambda =L_0+\lambda I_\beta,
\label{j12}
\end{equation}
where $I_\beta$ is an operator on $\h$ determined by a formal interaction Hamiltonian, such as those in \fer{j2} or \fer{j5}. The interaction $I_\beta$ has the property that the dynamics generated by $L_\lambda$ defines a $*$automorphism group $\{\sigma_{t,\lambda}\}$ of the von Neumann algebra $\mm_\beta\subset{\cal B}(\h)$ obtained by taking the weak closure of the algebra $\pi_\beta({\frak A})$. This means that, for every operator $A\in\mm_\beta$ and arbitrary $t\in\r$, the operator
\begin{equation}
\sigma_{t,\lambda}(A):=e^{itL_\lambda} A e^{-itL_\lambda}
\label{j13}
\end{equation}
belongs again to $\mm_\beta$. (For a representation-independent way of introducing interactions between the small system and the heat bath, see e.g. [FM].) Following ideas in [A], one can prove that, for a large class of interactions $I_\beta$, there exists a vector $\Obl\in\h$ with the property that the state
\begin{equation}
\obl(A):=\scalprod{\Obl}{A\Obl},\ \ \ A\in\mm_\beta
\label{j14}
\end{equation}
is an {\it equilibrium state} for the {\it interacting system}, in the sense that it satisfies the {\it Kubo-Martin-Schwinger (KMS) condition} for the interacting dynamics on the von Neumann algebra $\mm_\beta$, described by $\sigma_{t,\lambda}$; (see [HHW], or [JP2, BFS, DJP], for an explanation of these notions). The property of return to equilibrium means that the equilibrium state on $\mm_\beta$ given by $\obl$ is {\it dynamically stable}, in the sense of the following definition. \\

{\it Definition.\ } The system described by the von Neumann algebra $\mm_\beta$ and the time-evolution $\sigma_{t,\lambda}$ on $\mm_\beta$ (a so-called $W^*$-dynamical system) has the property of {\it return to equilibrium} iff, for an arbitrary normal state $\omega$ on $\mm_\beta$ (i.e., a state on $\mm_\beta$ given by a density matrix on $\h$) and an arbitrary operator $A\in\mm_\beta$,
\begin{equation}
\lim_{t\rightarrow\infty}\omega(\sigma_{t,\lambda}(A)) =\obl(A),
\label{j15}
\end{equation}
or (more modestly)
\begin{equation}
\lim_{t\rightarrow\infty}\frac{1}{t}\int_0^t ds \ \omega(\sigma_{s,\lambda}(A))= \obl(A),
\label{j16}
\end{equation}
(return to equilibrium in the sense of ergodic averages).\\

The convergence in \fer{j15} and \fer{j16} follows from the KMS condition for $\obl$ and certain spectral properties of the interacting Liouvillian, $L_\lambda$; see e.g. [JP2,BFS]. Because $\obl$ is invariant under the time evolution $\sigma_{t,\lambda}$, the interaction $\lambda I_\beta$ in \fer{j12} can be chosen s.t. 
\begin{equation}
L_\lambda\Obl=0,
\label{j17}
\end{equation}
i.e., zero is an eigenvalue of $L_\lambda$. If zero is a {\it simple} eigenvalue of $L_\lambda$ then, as a fairly easy consequence of the KMS condition and the von Neumann ergodic theorem, property \fer{j16} holds, and if the spectrum of $L_\lambda$ is absolutely continuous, except for a simple eigenvalue at zero, then \fer{j15} holds (this is again easily from the KMS condition and the RAGE theorem. Let us also mention that if the kernel of $L_\lambda$ is simple then $L_\lambda$ does not have any nonzero eigenvalues, see e.g. [JP3]). \\
\indent
The purpose of this paper is to exhibit a class of physically interesting interactions with the property that, for {\it all} $\beta$, with $(k_BT_0)^{-1}\equiv \beta_0<\beta<\infty$, return to equilibrium in the sense of ergodic averages, \fer{j16}, holds, provided the coupling constant $\lambda$ is small enough,
\begin{equation*}
0<|\lambda|<\lambda_0,
\end{equation*}
where, for infrared-regular interactions, $\lambda_0$ {\it only} depends on $\beta_0$, while, for infrared-singular interactions, $\lambda_0\rightarrow 0$ logarithmically, as $\beta\rightarrow \infty$; see Theorem \ref{rtethm}. This result relies, in part, on the following result: Given any $\epsilon>0$, there exists a positive constant $\lambda_1(\epsilon)$ and a choice of the phases of the vectors $\Obl$ and $\Obz$ such that 
\begin{equation}
\| \Obl-\Obz\| <\epsilon,
\label{j18}
\end{equation}
for all $\lambda$, with $|\lambda|<\lambda_1(\epsilon)$; in the infrared-regular case, the constant $\lambda_1(\epsilon)$ only depends on $\epsilon$, but is {\it independent} of $\beta$, and it decays to zero as $\beta\rightarrow\infty$ for infrared-singular systems; see Theorem \ref{betauniform}.\\

A proof of return to equilibrium in the stronger sense \fer{j15}, and {\it uniformly} in the temperature $0<T\leq T_0<\infty$ has been obtained already in [BFS] and in [DJ2], under the infrared conditions $g_0(k)\sim |k|^p$, ($|k|\sim 0$) for some $p>0$ and $p>2$, respectively. In addition, [DJ2] show \fer{j15} in the infrared-singular case \fer{j6}, for small coupling, tending to zero as $T\rightarrow 0$. The infrared conditions we impose to show \fer{j16} are $p=-1/2$ ($T$-dependent smallness of the coupling), and $p=1/2,3/2$, $p>2$ (small coupling, uniformly in $T$). \\

{\it Acknowledgements.\ } We thank V. Bach and I.M. Sigal for countless discussions on related problems and spectral methods without which this work would never have been done. M.M. is grateful to V. Jak\u si\'c for illuminating discussions. We have enjoyed the hospitality of IHES during initial and final stages of this work.

\subsection{The model}
\label{modelsection}
We consider a quantum system composed of a ``small'' subsystem interacting
with a ``large'' subsystem. The pure states of the small subsystem, which is also called {\it atom} (or
{\it spin}), are given by rays in the finite dimensional Hilbert space
\begin{equation}
\h_{at}=\cx^d.  
\label{100}
\end{equation}
The atomic Hamiltonian $H_{at}$ has simple eigenvalues $E_0<E_1 <\cdots < E_{d-1}$,  
\begin{equation}
H_{at}=\mbox{diag$(E_0,E_1,\ldots,E_{d-1})$}.
\label{101}
\end{equation}
It determines the dynamics $\alpha^{at}_t$ of observables $A\in{\cal B}(\h_{at})$ according to 
\begin{equation}
\alpha_t^{at}(A)=e^{itH_{at}} A e^{-itH_{at}}, 
\label{102}
\end{equation}
where $t\in\r$. For any inverse temperature
$0<\beta<\infty$ there is a unique $\beta$-KMS state on ${\cal B}(\h_{at})$
associated with the dynamics \fer{102}, called the atomic Gibbs state (at inverse temperature $\beta$). It is given by
\begin{equation}
\omega_\beta^{at}(\cdot)=\frac{\tr\left( e^{-\beta H_{at}}\ \cdot \right)}{\tr
  e^{-\beta H_{at}}},
\label{102'}
\end{equation}
where the trace is taken over $\h_{at}$. \\
\indent
The large subsystem is infinitely extended and is described by a free, scalar, 
massless Bose field. Its state is taken to be the equilibrium state at inverse temperature $0<\beta<\infty$. The description of this state and the GNS representation is standard (see e.g. [AW], [JP1,2], [FM]). We present only the essentials and point out a modification we introduce (namely the phase $\phi$ in \fer{114}). 
Let \begin{equation}
L^2_0:=L^2(\r^3,d^3k)\cap L^2(\r^3,|k|^{-1}d^3k)
\label{103}
\end{equation}
and denote by ${\frak W}(L^2_0)$ the Weyl algebra over $L^2_0$, i.e., the
$C^*$-algebra generated by Weyl operators $W(f)$, $f\in L^2_0$, satisfying the
CCR 
\begin{equation}
W(f)W(g)=e^{-\frac{i}{2}{\rm Im}\scalprod{f}{g}} W(f+g) =
e^{-i{\rm Im}\scalprod{f}{g}} W(g)W(f),
\label{104}
\end{equation}
and the relations $W(f)^*=W(-f)$, $W(0)=\bbbone$ (unitarity). The brackets $\scalprod{\cdot}{\cdot}$ in \fer{104} denote the inner product of $L^2(\r^3,d^3k)$.
The large subsystem is described by the $\beta$-KMS state $\omega_\beta^f$ on 
${\frak W}(L^2_0)$ associated with the dynamics
\begin{equation}
\alpha_t^f(W(f))=W(e^{it\omega}f),
\label{105}
\end{equation}
with dispersion relation  
\begin{equation}
\omega(k)=|k|.
\label{106}
\end{equation}
\indent
An interaction between the two subsystems can be specified in a representation independent way in terms of a
suitable $*$automorphism group $\alpha_{t,\lambda}$ on the $C^*$-algebra
${\cal B}(\h_{at})\otimes {\frak W}(L^2_0)$, where $\lambda$ is a perturbation
parameter and $\alpha_{t,0}=\alpha_t^{at}\otimes \alpha_t^f$. Here we do not
discuss this procedure of defining $\alpha_{t,\lambda}$ -- this has been
discussed in [FM]. Rather, we directly specify how the interacting dynamics acts (is implemented) on the GNS
Hilbert space corresponding to 
\begin{equation}
\obz=\omega_\beta^{at}\otimes \omega_\beta^f,
\label{107}
\end{equation}
the $(\beta, \alpha_{t,0})$-KMS state on the algebra ${\frak A}={\cal B}(\h_{at})\otimes
{\frak W}(L^2_0)$.
The GNS representation of the algebra $\frak A$ determined by the state \fer{107} is explicitly given in 
[AW] and has been put, in [JP1,2], in a form adapted to the use of the theory of spectral deformations (and of positive commutators). We use a slight modification of the representation in [JP1,2]. 
The representation Hilbert space is 
\begin{equation}
\h=\h_{at}\otimes\h_{at}\otimes {\cal F},
\label{108}
\end{equation}
where 
\begin{equation}
{\cal F}={\cal F}\left(L^2(\r\times S^2, du\times d\sigma)\right)
\label{109}
\end{equation}
is the bosonic Fock space over $L^2(\r\times S^2, du\times d\sigma)$, where $d\sigma$ denotes the uniform measure on $S^2$. We use the following notational convention: we write $L^2(\r\times S^2)$ for $L^2(\r\times S^2, du\times d\sigma)$ and $L^2(\r^3)$ stands for $L^2(\r^3,d^3k)$, or for $L^2(\r_+\times S^2, u^2du\times d\sigma)$ (polar coordinates). \\
\indent
The cyclic vector representing $\obz$ in $\h$ is
\begin{equation}
\Obz=\Omega_\beta^{at}\otimes \Omega.
\label{110}
\end{equation}
Here $\Omega$ is the vacuum vector in $\cal F$ and
\begin{equation}
\Omega_\beta^{at} =\left(\tr e^{-\beta H_{at}}\right)^{-1} \sum_{j=0}^{d-1}
e^{-\beta E_j/2}\varphi_j\otimes \varphi_j,
\label{111}
\end{equation}
where $\varphi_j$ is the eigenvector of $H_{at}$ associated to the eigenvalue
$E_j$, see also \fer{101}. To complete our description of the GNS
representation of \fer{107} we need to give the representation map $\pi_\beta:
{\cal B}(\h_{at})\otimes {\frak W}(L^2_0)\rightarrow {\cal B}(\h)$. It is the
product
\begin{equation}
\pi_\beta= \pi^{at}\otimes \pi_\beta^f,
\label{112}
\end{equation}
with
\begin{eqnarray}
\pi^{at}(A)&=& A\otimes \bbbone_{at}\label{113'}\\
\pi_\beta^f(W(f))&=& e^{i\varphi(\tau_\beta f)},
\label{114'}
\end{eqnarray}
and where, for $h\in L^2(\r\times S^2)$, $\varphi(h)$ is the
selfadjoint operator on $\cal F$ given by 
\begin{equation}
\varphi(h)=\frac{a^*(h)+a(h)}{\sqrt 2}.
\label{113}
\end{equation}
The operators $a^*(h)$ and $a(h)$ are standard creation and annihilation operators on $\cal F$, smeared out with $h$. We take $h\mapsto a^*(h)$ to be linear. The real-linear map $\tau_\beta: L^2_0\rightarrow L^2(\r\times S^2)$ appearing in \fer{114'} acts as 
\begin{equation}
(\tau_\beta f)(u,\sigma) =\sqrt{\frac{u}{1-e^{-\beta u}}} \left\{
\begin{array}{ll}
\sqrt{u} \ f(u,\sigma), & u>0,\\
\sqrt{-u}\  e^{i\phi} \overline{f}(-u,\sigma), & u<0,
\end{array}
\right.
\label{114}
\end{equation}
where we represent $f$ in polar coordinates and $\overline f$ means the complex conjugate of $f$. We have introduced an arbitrary phase $\phi\in\r$ which can
be chosen conveniently so as to tune discontinuity properties of the r.h.s. in \fer{114} at $u=0$. 
The origin of this freedom can be explained as follows. 
The expectation functional of $\omega_\beta^f$ is given by
\begin{equation}
L^2_0\ni f\mapsto \omega_\beta^f(W(f))= \exp\left[ -\frac{1}{4}\int_{\r^3}\left( 1+\frac{2}{e^{\beta|k|}-1}\right) |f(k)|^2 d^3k\right],
\label{114.1}
\end{equation}
which corresponds to the state of black body radiation at inverse temperature $\beta$, see [AW]. We define a family of (equivalent) representations of the Weyl algebra ${\frak W}(L^2_0)$ on the Hilbert space \fer{109} by the map
\begin{equation}
\pi_\beta^{U_+,U_-}\big(W(f)\big) =\exp\left[i\varphi\left(\tau_\beta^{U_+,U_-}f\right)\right],
\label{114.2}
\end{equation}
where $\varphi$ is defined in \fer{113}, $U_+$, $U_-$ are arbitrary unitary operators on $L^2(\r^3)$, and 
\begin{equation}
\left(\tau_\beta^{U_+,U_-}f\right)(u,\sigma)=
\left\{
\begin{array}{ll}
u (U_+(1-e^{-\beta u})^{-1/2}f)(u,\sigma), & u>0,\\
u (U_-(e^{-\beta u}-1)^{-1/2}\overline{f})(-u,\sigma), & u<0.
\end{array}
\right.
\label{114.3}
\end{equation}
It is easily seen that, for any choice of the unitaries $U_\pm$, 
\begin{equation*}
\scalprod{\Omega}{\exp\left[i\varphi\left(\tau_\beta^{U_+,U_-}f\right)\right] \Omega}
\end{equation*}
equals the r.h.s. of \fer{114.1}. Expression \fer{114.3} reduces to \fer{114} for $U_+=id$, $U_-=e^{i\phi}$.\\
\indent
{\it Remark.\ } We recall that there is a second representation, $\widetilde\pi_\beta^{U_+,U_-}$ of ${\frak W}(L^2_0)$ on $\cal F$ given by
\begin{equation}
\widetilde\pi_\beta^{U_+,U_-}\big(W(f)\big) =\exp\left[ i\varphi\left(\tau_\beta^{U_+,U_-}\left(e^{-\beta u/2}f\right)\right)\right],
\label{114.4}
\end{equation}
which commutes with the representation $\pi_\beta^{U_+,U_-}$. \\
\indent
In previous articles involving this setting, [JP1,2], [BFS], [DJ1], [M], [DJ2],
[FM], [FMS], the freedom of choosing $U_\pm$ arbitrarily was not used, only $U_\pm=\pm id$ was considered. For a suitable choice of $U_\pm$ one can apply the existing positive commutator methods, based on the generator of translations in $u\in\r$ as conjugate
operator, to models with fermionic or bosonic fields having dispersion relation different from \fer{106}. These matters will be pursued in another work. Here we restrict our attention to the representation \fer{114}, where $\phi$ is a phase determined by the interaction, see assumption (A1) and the discussion thereafter.\\
\indent
We are now ready to define the interacting dynamics as the  
$*$automorphism group 
\begin{equation}
\sigma_{t,\lambda}(\cdot)=e^{itL_\lambda}( \cdot )e^{-itL_\lambda}
\label{dyn}
\end{equation}
on the von Neumann algebra
\begin{equation}
\mm_\beta:= \pi_\beta\left({\cal B}(\h_{at})\otimes {\frak
    W}(L^2_0)\right)''\subset {\cal B}(\h),
\label{116}
\end{equation}
where $''$ denotes the double commutant (weak closure), and where 
the generator $L_\lambda$, called the {\it standard Liouvillian} of the
system, is the selfadjoint operator on $\h$ given by ([JP1,2], [FM])
\begin{equation}
L_\lambda= L_0+\lambda I,
\label{117}
\end{equation}
with
\begin{equation}
L_0=L_{at}+L_f, \ \ \ \ \ L_{at}=H_{at}\otimes \bbbone_{at} -\bbbone_{at}\otimes H_{at},  \ \ \
\ \  L_f=\d\Gamma(u). 
\label{118}
\end{equation}
Here, $\d\Gamma(u)$ denotes the second quantization (acting on $\cal F$) of
the operator of multiplication by $u\in\r$, $\lambda$ is a coupling constant,
and $I$ is the finite sum
\begin{equation}
I= \sum_\alpha \Big\{ G_\alpha\otimes\bbbone_{at}\otimes
\varphi(\tau_\beta(g_\alpha)) 
-\bbbone_{at}\otimes \cc_{at} G_\alpha \cc_{at}\otimes\varphi
(\tau_\beta(e^{-\beta u/2}g_\alpha))\Big\},
\label{119}
\end{equation}
where the operators $G_\alpha$ are bounded, selfadjoint operators on $\h_{at}$, and
the functions $g_\alpha\in L^2_0$ are called {\it form factors}. $\cc_{at}$ is
the antilinear operator of component-wise
complex conjugation in the basis $\{\varphi_j\}_{j=0}^{d-1}$ diagonalizing
$H_{at}$. Note that $L_0$ does not depend on the choice of the phases $\phi_\pm$,
but $I$ does.
The following relative bounds are standard
\begin{equation}
\| I(N+1)^{-1/2}\|,\  \|(N+1)^{-1/2}I\| < C(1+1/\beta),
\label{standardestimate}
\end{equation}
where $C$ is some constant which is independent of $\beta$. \\
\indent
At temperature zero ($\beta=\infty$), the Liouvillian
\fer{117} corresponds to the Hamiltonian  
\begin{equation}
H_\lambda= H_{at} +\d\Gamma(\omega) +\lambda\sum_\alpha G_\alpha\otimes
\varphi(g_\alpha),
\label{hamiltonian}
\end{equation}
which describes interactions of the atom with the quantized field involving emission and absorption of field quanta.\\
\indent
The pair $(\mm_\beta,\sigma_{t,\lambda})$ is called a $W^*$-dynamical system. For $\lambda=0$ the state on $\mm_\beta$ determined by $\Obz$ is a $(\beta,\sigma_{t,0})$-KMS state. It is well known ([A], [FNV], [BFS], [DJP]) that the vector
\begin{equation}
\Obl:=Z^{-1}_{\beta,\lambda}\ e^{-\beta(L_0+\lambda I_\ell)/2} \Obz \in \h,
\label{Obl}
\end{equation}
where $Z_{\beta,\lambda}$ is a normalization factor, and $I_\ell$ is obtained from $I$ by dropping the second term in the sum \fer{119}, defines a $(\beta, \sigma_{t,\lambda})$-KMS state on $\mm_\beta$.\\
\indent Before stating our results we make two assumptions on the interaction.
\begin{itemize}
\item[(A1)] 
The form factors are given by $g_\alpha(u,\sigma) = u^p\widetilde g_\alpha(u,\sigma)$, where $p$ takes one of the values $-1/2, 1/2, 3/2$ or
  $p>2$, and the $\widetilde g_\alpha$ satisfy a set of conditions we describe next. For fixed $\sigma$ and $\alpha$, the map $u\mapsto
  \widetilde g_\alpha(u,\sigma)$ is $C^3$ on $(0,\infty)$ and 
\begin{equation}
\|\partial_u^j\widetilde g_\alpha\|_{L^2(\r^3)}<\infty,\mbox{\ \ for
  $j=0,1,2,3$}.   \label{120}
\end{equation}
If $p=-1/2, 1/2$ or $3/2$ then the limits 
\begin{equation}
\partial_u^j \widetilde g_\alpha(0,\sigma):=\lim_{u\rightarrow 0_+} \partial_u^j\widetilde
g_\alpha(u,\sigma)
\label{121}
\end{equation}
exist, for $j=0,1,2$, and there is a phase $\phi_0\in\r$, not depending on $\alpha, \sigma$ and $j=0,1,2$, s.t.
\begin{equation}
e^{-i\phi_0}\partial_u^j\widetilde g_\alpha(0,\sigma) \in \r.
\label{121.1}
\end{equation}
In addition, if $p=-1/2, 1/2$ then we require 
$\partial_u \widetilde g_\alpha(0,\sigma)=0$. Finally, we assume that 
\begin{equation}
\|u^2 g_\alpha\|_{L^2(\r^3)}<\infty.
\label{122}
\end{equation}
\item[(A2)] It is assumed that 
\begin{equation}
\min_{E_m\neq E_n} \int_{S^2}d\sigma \left| \sum_\alpha
  \scalprod{\varphi_m}{G_\alpha \varphi_n}
  g_\alpha\left(|E_m-E_n|,\sigma\right)\right|^2 >0.
\label{123}
\end{equation}
\end{itemize}
{\it Discussion of assumptions (A1) and (A2).\ }
Assumption (A1) concerns smoothness and decay properties of the form
factors, which are necessary in the application of the Virial Theorem, see the remark after Theorem \ref{vthm}. If the interaction is characterized, according to (A1), by $p=-1/2$, then we choose the phase $\phi$ in \fer{114} to be $\phi=2\phi_0$. For all other values of $p$ we take $\phi=\pi+2\phi_0$. For $p=-1/2,1/2$ an admissible infrared behaviour of the form factors is $g_\alpha\sim u^p$ times a constant, as $u\sim 0$. Other than for the applicability of the Virial Theorem, condition \fer{122} is also used to show that $L_\lambda$ is selfadjoint (for any
$\lambda\in \r$). This follows from the Glimm-Jaffe-Nelson commutator theorem,
see [FM]. \\
\indent
Assumption (A2) is called the {\it Fermi Golden Rule Condition} and has been
discussed extensively in previous works, see e.g. [JP1,2], [BFS], [DJ1], [M], [DJ2], [DJ3]. Its role is to guarantee that the probability of absorption and emission processes of field quanta does not vanish in second order perturbation theory (in $\lambda$). This can be translated into a suitable positivity condition on an operator $\Gamma_0$, called the {\it level shift operator}, see \fer{127} below. Let 
\begin{equation}
\Pi= P_0\otimes P_\Omega
\label{125'}
\end{equation} 
denote the projection onto the kernel of $L_0$, where $P_0$ is the
rank-$d$ projection onto the kernel of $L_{at}$, and $P_\Omega$ is the
projection onto $\cx \Omega$, $\Omega$ being the vacuum vector in $\cal F$, see
\fer{109}. We will see that if the non-negative operator $\Pi
I\delta(L_0)I\Pi$, where $\delta$ is the Dirac distribution, has a
one-dimensional kernel (the dimension is at least one, since the kernel contains the
atomic Gibbs state \fer{111}) then the system has the property of return to
equilibrium.
\begin{theorem}
Assume \fer{123}. There is an
$\epsilon_0>0$, independent of $\beta\geq\beta_0$ (for any $\beta_0$
fixed), s.t. if $0<\epsilon<\epsilon_0$ then 
\begin{equation}
\Pi I\frac{\epsilon}{L_0^2+\epsilon^2} I\Pi \geq \Gamma_0\Pi -C\epsilon^{1/4},
\label{126}
\end{equation}
where $C$ is a constant independent of $\beta$, and $\Gamma_0$ is a bounded operator on $\h=\h_{at}\otimes\h_{at}\otimes{\cal F}$, acting trivially on the last factor, $\cal F$, and leaving $\ker L_{at}$ invariant. Moreover, $\Gamma_0$ restricted to $\ker L_{at}$ has zero as a simple eigenvalue, with the atomic Gibbs state
  $\Omega_\beta^{at}$ as eigenvector, see \fer{111}, and is strictly positive on
the complement of $\cx\Omega_\beta^{at}$. More precisely, 
there is a constant
$\gamma_0>0$, independent of $0<\beta<\infty$, s.t.
\begin{equation}
\Gamma_0\upharpoonright_{\ran \Pbar_{\Omega_\beta^{at}}}\geq \gamma_0.
\label{127}
\end{equation}
Here, $\Pbar_{\Omega_\beta^{at}}=\bbbone -P_{\Omega_\beta^{at}}$ and
$P_{\Omega_\beta^{at}}$ is the projection onto $\cx \Omega_\beta^{at}$. 
\end{theorem}
A proof of this result, in the case where the sum in \fer{119} reduces to a
single 
term, can be in found [M], [BFS]. It is easy to carry out that proof for the more general 
interaction \fer{119}. An explicit lower bound, in terms of \fer{123}, can be given: 
\begin{eqnarray*}
\lefteqn{
\Gamma_0\upharpoonright_{\ran \Pbar_{\Omega_\beta^{at}}}}\\
&&\geq
\min_{E_m\neq E_n} \frac{(E_m-E_m)^2\ \tr e^{-\beta H_{at}} }{|e^{-\beta E_m}-e^{-\beta E_n}|}
\int_{S^2}d\sigma \left| \sum_\alpha
  \scalprod{\varphi_m}{G_\alpha \varphi_n}
  g_\alpha\left(|E_m-E_n|,\sigma\right)\right|^2,
\end{eqnarray*}
which yields $\gamma_0$ in \fer{127} by minimizing the r.h.s. over $0<\beta<\infty$.

\section{Main results}
\label{mainresultsection}
Our main result on return to equilibrium is 
\begin{theorem}{\bf (Return to equilibrium).\ }
\label{rtethm}
Assume conditions (A1) and (A2). There is a constant $\lambda_0>0$, independent of $\beta\geq \beta_0$, for any $\beta_0>0$ fixed, s.t. if 
\begin{equation}
0<|\lambda| <\lambda_0\left\{
\begin{array}{cl}
1 &\mbox{if $p>-1/2$}\\
\big(1+\log(1+\beta)\big)^{-9/2} & \mbox{if $p=-1/2$},
\end{array}
\right.
\label{f4}
\end{equation}
 then the kernel of $L_\lambda$ is spanned by the interacting KMS vector $\Obl$, \fer{Obl}. In other words, the system has the property of return to equilibrium. 
\end{theorem}
If the temperature of the heat bath is very large then second order processes of absorption and emission of field quanta do not dominate the ones of higher order, and we cannot expect to describe the physics of the system using perturbation theory in second order (although, for the {\it toy atom} considered here, the phenomenon of return to equilibrium is expected to take place at {\it all} temperatures; see also our discussion in the introduction). 
This is why, in the following analysis, the dependence of the constant $\lambda_0$ in Theorem \ref{rtethm} on $\beta_0$ is natural. 
The fact that, for $p=-1/2$, we must impose an upper bound on the coupling constant tending to zero, as $\beta\rightarrow\infty$ (see \fer{f4}), can be understood as follows: Our methods are perturbative (in $\lambda$) and rely on controlling the (norm-) distance between the KMS states for the interacting and the non-interacting systems (see Theorem \ref{betauniform}). One cannot, in general, expect this distance to be small, for small but non-zero coupling constants, uniformly in $\beta\rightarrow \infty$. 
This is due to the fact that, for $p=-1/2$, and in the zero temperature limit, $\beta\rightarrow\infty$, the groundstate of an interacting, infrared singular system is {\it not} in Fock space (i.e., the Hamiltonian \fer{hamiltonian} does not have a groundstate in $\h_{at}\otimes{\cal F}(L^2(\r^3))$, see e.g. [AH], [Sp]), but the non-interacting system ($\lambda=0$) does have a groundstate in Fock space! Consequently, we expect the difference between the interacting and the non-interacting KMS state to diverge, as $\beta\rightarrow\infty$, for $p=-1/2$.\\
\indent
Assuming that the interaction between the small system and the heat bath is such that $\scalprod{\Obl}{N\Obl}$ is small, for small values of $\lambda$, uniformly in $\beta\rightarrow\infty$, where $N=\d\Gamma(\bbbone)$ is the number operator in the Araki-Woods representation and $\Obl$ is given in \fer{Obl}, then our methods can be used to establish return to equilibrium  for sufficiently small values of $\lambda$, {\it uniformly} in $\beta\geq \beta_0$, even when $p=-1/2$.\\
\indent
From a more technical point of view, we can describe the above discussion as follows. A typical estimate involved in our analysis is inequality \fer{standardestimate}, where $C$ is some finite constant. Then $\|\lambda I (N+1)^{-1/2}\|$ can be made smaller than any constant $\delta>0$, provided $|\lambda|$ is chosen sufficiently small, {\it independently} of $\beta>\beta_0$, for an arbitrary, but fixed $\beta_0>0$. Similarly, in order to estimate the norm of the difference between the interacting and the non-interacting KMS state, we need an upper bound on the expectation value of the number operator $N$ in the interacting KMS state $\Obl$.
As explained  after the statement of Theorem \ref{betauniform}, this expectation value 
 is bounded above by $\|\lambda I_1(N+1)^{-1/2}\|$, where $I_1$ is defined in \fer{204}. For $p=-1/2$, the latter norm is not uniformly bounded in $\beta\geq\beta_0$, but diverges logarithmically, as $\beta\rightarrow\infty$. Thus, requiring it to be bounded by a small constant, we must assume that $|\lambda| \log(\beta)$ is sufficiently small. \\
\indent
Among the technical results used in our proof of Theorem \ref{rtethm} we single out the following one, which shows that the perturbed and unperturbed KMS states are close to each other, for small coupling constants. In the infrared-regular regime $p>-1/2$, the difference between the two KMS states is small {\it independently} of the inverse temperature.
\begin{theorem}
\label{betauniform}
Assume (A1) and let $P_\Obl$ and $P_\Obz$ denote the projections onto the spans of the interacting and non-interacting KMS states, $\Obl$ (see \fer{Obl}) and $\Obz$ (see \fer{110}), respectively. For any $\epsilon>0$ there is a $\lambda_0(\epsilon)>0$, which does not depend on $\beta>0$, s.t. if 
\begin{equation}
|\lambda|<\lambda_0(\epsilon) \left\{
\begin{array}{cl}
1 &\mbox{if $p>-1/2$}\\
\big(1+\log(1+\beta)\big)^{-1} &\mbox{if $p=-1/2$}
\end{array}
\right.
\label{f1}
\end{equation}
then
\begin{equation}
\left\| P_\Obl -P_\Obz\right\|<\epsilon.
\label{1}
\end{equation}
\end{theorem}
{\it Remark.\ } 
The constant $\lambda_0(\epsilon)$ in Theorem \ref{betauniform} depends on the spectral gap $E_1-E_0>0$ of the atomic Hamiltonian, and, if the norms $\|G_\alpha\|$ are assumed to satisfy a $d$-independent upper bound, then $\lambda_0(\epsilon)$ can be chosen independently of the dimension $d$ of the atomic Hilbert space. 

We prove Theorem \ref{betauniform} in Section \ref{betauniformsect}.

\section{Proof of Theorem \ref{rtethm}}
\label{rtethmproof}
We use a simplified version of the {\it positive commutator (PC) method}, introduced in the present context, for zero temperature systems, in [BFSS], and extended to the positive temperature situation in [M]. We refer to [DJ1], [DJ2], [O], [GGS1], [GGS2], and to the book [ABG], for recent different implementations of this method.

\subsection{Mechanism of the proof}
There are two key ingredients in our proof, the PC estimate and the Virial Theorem. While we give a proof of the PC estimate, we refer to [FM] for a proof of the Virial Theorem.  \\
\indent
Let $A_f=\d\Gamma(i\partial_u)$ be the second quantization of $i\partial_u$ on $\cal F$ (c.f. \fer{109}) and set
\begin{equation}
A_0=i\theta\lambda\left( \Pi I\repsilonbar^2-\repsilonbar^2 I\Pi\right),
\label{201}
\end{equation}
where $\Pi=P_0\otimes P_\Omega$ has been defined in \fer{125'}, $\repsilonbar=\Pibar\repsilon$, $\Pibar=\bbbone-\Pi$, $\repsilon =(L_0^2+\epsilon^2)^{-1/2}$, and $\theta,\epsilon$ are positive parameters. We note that $A_0$ is a bounded operator satisfying $\ran A_0\subset\dom (L_\lambda)$, and that the commutator $[L,A_0]$ extends to a bounded operator with
\begin{equation}
\left\|[L_\lambda,A_0]\right\|\leq C\left(\frac{\theta|\lambda|}{\epsilon}+\frac{ \theta\lambda^2}{\epsilon^2}\right).
\label{202}
\end{equation}
On the domain $\dom(N)$ of the number operator $N=\d\Gamma(\bbbone)$ we define the operator 
\begin{equation}
B=N+\lambda I_1+i[L_\lambda,A_0],
\label{203}
\end{equation}
where
\begin{equation}
I_1= \sum_\alpha \big( G_\alpha\otimes\bbbone_{at}\otimes
\varphi(\partial_u\tau_\beta(g_\alpha)) 
-\bbbone_{at}\otimes \cc_{at} G_\alpha \cc_{at}\otimes\varphi
(\partial_ue^{-\beta u/2}\tau_\beta(g_\alpha))\big).
\label{204}
\end{equation}
The operator $B$ represents the quadratic form $i[L_\lambda,A_f+A_0]$, see [FM]. 
\begin{theorem}{\rm (\bf Positive commutator estimate).} 
\label{pcthm}
Assume (A1) and (A2), and fix $0<\eta<2/3$. For any $\nu>1$ set 
\begin{equation*}
{\frak B}_\nu=\{\psi\in \dom(N^{1/2})\ |\ \|\psi\|=1, \|(N+1)^{1/2}\psi\|\leq\nu\}.
\end{equation*}
There is a choice of the parameters $\epsilon$ and $\theta$, and a constant $\lambda_1(\eta)=\lambda_1>0$, not depending on $\nu$ and $\beta\geq\beta_0$, s.t. if
\begin{equation}
0<|\lambda|<\lambda_1 \left\{
\begin{array}{cl}
1&\mbox{if $p>-1/2$}\\
\min\left(\frac{1}{1+\log(1+\beta)}, \frac{\nu^{1/\eta-9/2}}{(1+\log(1+\beta))^\eta}\right) &\mbox{if $p=-1/2$},
\end{array}
\right.
\label{f3}
\end{equation}
then we have
\begin{equation}
\Pbar_\Obl B\Pbar_\Obl \geq |\lambda|^{2-\eta}\nu^{3-9\eta/2} \gamma_0\Pbar_\Obl,
\label{pc}
\end{equation}
in the sense of quadratic forms on $\ran E_\Delta(L_\lambda)\cap{\frak B}_\nu$, where $\Delta$ is any interval around the origin s.t. $\Delta \cap\sigma(L_{at})=\{0\}$, $E_\Delta(L_\lambda)$ is the spectral projection, and
where $\gamma_0$ is given in \fer{127}. 
\end{theorem}
We note that it is enough, for our purposes, to examine $B$ as a quadratic form on a subset of $\dom(N^{1/2})$, because any eigenvector $\psi_\lambda$ of $L_\lambda$ satisfies $\psi_\lambda\in{\frak B}_{\nu_0}$, for some $\nu_0$ which is independent of $|\lambda|\leq 1$. Moreover, for $p>-1/2$, $\nu_0$ does not depend on $\beta\geq\beta_0$, while for $p=-1/2$, $\nu_0$ diverges logarithmically for large $\beta$. These facts follow from the next result.
\begin{theorem}{\bf (Regularity of eigenvectors and Virial Theorem, [FM], [FMS]).\ }
\label{vthm}
Assume (A1). Let $\psi_\lambda$ be an eigenvector of $L_\lambda$. There is a constant $c(p,\beta)<\infty$, not depending on $\lambda$, s.t.
\begin{equation}
\|N^{1/2}\psi_\lambda\|\leq c(p,\beta) |\lambda|\, \|\psi_\lambda\|,
\label{205}
\end{equation}
and s.t. for all $\beta\geq \beta_0$ (for any $\beta_0>0$ fixed),
\begin{eqnarray}
c(p,\beta)\leq c_1(p)
\left\{
\begin{array}{cl}
1 & \mbox{if $p>-1/2$}\\
1+\log(1+\beta) &\mbox{if $p=-1/2$}
\end{array}
\right.
\label{cpb1}
\end{eqnarray}
where $c_1$ does not depend on $\beta\geq \beta_0$. Moreover, 
\begin{equation}
\av{B}_{\psi_\lambda}:=\scalprod{\psi_\lambda}{B\psi_\lambda}=0.
\label{206}
\end{equation}
\end{theorem}
{\it Remarks.\ } The constant $c(p,\beta)$ can be expressed in terms of the operator $I_1$ given in \fer{204} as follows: 
\begin{equation*}
\| I_1(N+1)^{-1/2}\|\leq 2\sum_\alpha \|G_\alpha\|\ \|\partial_u\tau_\beta(g_\alpha)\|_{L^2}=c(p,\beta).
\end{equation*}
One can understand \fer{206} formally by expanding the commutator, 
\begin{equation}
\scalprod{\psi_\lambda}{[L_\lambda,A_f+A_0] \psi_\lambda} =2 i \IM \scalprod{L_\lambda\psi_\lambda}{(A_f+A_0)\psi_\lambda}=0.
\label{206.1}
\end{equation}
The same argument gives $\av{[L_\lambda,A_f]}_{\psi_\lambda}=0$, from which it follows that 
\begin{eqnarray}
0\geq \av{N}_{\psi_\lambda} -\left| \av{\lambda I_1}_{\psi_\lambda}\right|&\geq& \av{N}_{\psi_\lambda} -c(p,\beta)|\lambda|\, \|\psi_\lambda\|\, \|N^{1/2}\psi_\lambda\| \nonumber\\
&\geq& \frac{1}{2} \av{N}_{\psi_\lambda}-\frac{1}{2}c(p,\beta)^2\lambda^2\|\psi_\lambda\|^2,
\label{206.2}
\end{eqnarray}
which yields the bound \fer{205}. 
In order to make the arguments leading to \fer{206} rigorous, one needs to control multiple commutators of $L_\lambda$ with $A_f+A_0$ of order up to three. 
In particular, we need the first, second and third commutator of $I$ with the dilation
generator $A_f$ to be a well-defined, relatively $N^{1/2}$-bounded operator, see [M], [FM]. 
The latter condition is satisfied provided
\begin{eqnarray}
\partial_u^j\tau_\beta(g_\alpha) \mbox{\ \ is continuous in $u\in\r$ for $j=0,1,2$,
  and }\label{124}\\ 
\partial_u^j\tau_\beta(g_\alpha)\in L^2(\r\times
  S^2) \mbox{\ \ for $j=0,1,2,3$}.\ \ \ \ \ \ \ \label{125}
\end{eqnarray}
We point out that for this argument, i.e. for the proof of \fer{206}, the $L^2$-norms of the functions $\partial_u^j\tau_\beta(g_\alpha)$ do not need to be bounded uniformly in $\beta$. 
It is not difficult to verify that \fer{124}, \fer{125} follow from (A1). Let $p$ and $\phi_0$ be as in assumption (A1); then, for
$p=1/2, 3/2$, $p>2$, we use the representation \fer{114} with $\phi=2\phi_0$, while for $p=-1/2$, we take $\phi=\pi+2\phi_0$.\\

The proof of Theorem \ref{rtethm} is an easy consequence of Theorems \ref{pcthm} and \ref{vthm}. Indeed, if, for $\lambda$ satisfying \fer{f3}, with $\nu=\nu_0$  (introduced after Theorem \ref{rtethm}), there were a unit eigenvector $\psi_\lambda\in\ker L_\lambda$, orthogonal to $\Obl$, then
\begin{equation}
0=\av{B}_{\psi_\lambda}\geq |\lambda|^{2-\eta}\nu_0^{3-9\eta/2}\gamma_0.
\label{207}
\end{equation}
Relation \fer{207} cannot hold since the r.h.s. is strictly positive. For $p=-1/2$ condition \fer{f3} (with $\nu=\nu_0=C[1+\log(1+\beta)]$) gives \fer{f4}, independently of $\eta$.

\subsection{Proof of Theorem \ref{pcthm}}
Since $\Omega_{\beta,\lambda}$ is in the kernel of $L_\lambda$, the commutator $B$ given in \fer{203} cannot be strictly positive on the entire space; see \fer{206}. To show that $\dim\ker L_\lambda=1$ it is
natural to try to show that 
\begin{equation}
B+\delta P_\Obl\geq \gamma,
\label{61}
\end{equation}
for some $\delta\geq \gamma$, where $\gamma>0$. Let $\Delta\subset \r$ be
an interval around the origin not containing any non-zero eigenvalue of the
atomic Liouvillian $L_{at}$. In Subsection \ref{ss1} we prove \fer{61} in the sense of quadratic forms on the 
spectral subspace of $L_0$ associated with the 
interval $\Delta$ (see \fer{74}). Using this inequality, we show in Subsection \ref{ss2} that
\begin{equation}
\Pbar_\Obl B\Pbar_\Obl \geq \frac{1}{2}\gamma\Pbar_\Obl,
\label{61.1}
\end{equation}
in the sense of quadratic forms on $\ran E_{\Delta'}(L_\lambda)\cap {\frak B}_\nu$, where $E_{\Delta'}(L_\lambda)$ is the spectral projection of $L_\lambda$ associated to an interval $\Delta'$, which can be chosen arbitrarily, as long as it is properly contained in $\Delta$.

\subsubsection{PC estimate localized w.r.t. $L_0$}
\label{ss1}
We will use the Feshbach method with the decomposition 
\begin{equation}
\h_\Delta^0:=\ran E_\Delta^0=\ran E_\Delta^0\Pi\oplus \ran E_\Delta^0\Pibar,
\label{62}
\end{equation}
where $\Pi$ is given in \fer{125'}, and 
where $E_\Delta^0$ is the spectral projection of $L_0$ associated with the
interval $\Delta$. For a presentation of this method resembling most closely the form in which it is used here we refer to [M], [FM], and, for more background, to [BFSS], [BFS], [DJ].\\
\indent
In what follows, $C$ denotes a constant independent of $\lambda, \theta,\epsilon, \beta\geq\beta_0$ (for any fixed $\beta_0>0$), and $C(p,\beta)$ denotes a constant independent of $\lambda,\theta,\epsilon$, satisfying the bound given in \fer{cpb1}. The values of $C$, $C(p,\beta)$ can vary from expression to expression.\\
\indent
{}From $\Pibar=\Pbar_0\otimes P_\Omega +\Pbar_\Omega$ and the properties of $\Delta$ it
follows that $\ran E_\Delta^0\Pibar\subset \ran \Pbar_\Omega$ and
\begin{eqnarray}
E_\Delta^0\Pibar(B+\delta P_\Obl)\Pibar E_\Delta^0&=& E_\Delta^0\Pibar
N^{1/2}\left( \bbbone +N^{-1/2}\lambda I_1N^{-1/2}\right) N^{1/2}\Pibar
E_\Delta^0\nonumber\\
&& + \edeltanot\Pibar \left( i[L_\lambda,A_0] +\delta P_\Obl\right)
\Pibar\edeltanot\nonumber\\
&\geq& \frac{1}{2}\edeltanot \Pibar +\edeltanot \Pibar
i[L_\lambda,A_0]\Pibar\edeltanot\nonumber\\
&\geq& \frac{1}{2}\left(1-C\frac{\theta\lambda^2}{\epsilon^2}
\right)\edeltanot\Pibar, 
\label{63}
\end{eqnarray}
provided
\begin{equation}
\|\Pbar_\Omega N^{-1/2}\lambda I_1 N^{-1/2}\Pbar_\Omega\|\leq C(p,\beta) |\lambda| <1/2,
\label{parcond1}
\end{equation}
see the remark after Theorem \ref{vthm}, and where we use the bound 
\begin{equation}
\|\edeltanot\Pibar [L_\lambda,A_0]\Pibar\edeltanot\| \leq
C\frac{\theta\lambda^2}{\epsilon^2}
\label{64}
\end{equation}
which follows easily from the definition of $A_0$, \fer{201}. We choose the
parameters s.t.
\begin{equation}
C\frac{\theta\lambda^2}{\epsilon^2}<1/2,
\label{parcond2}
\end{equation}
and hence we have that
\begin{equation}
\edeltanot\Pibar(B+\delta
P_\Obl)\Pibar\edeltanot\geq\frac{1}{4}\edeltanot\Pibar. 
\label{65}
\end{equation}
The Feshbach map associated with the decomposition \fer{62}
and with the spectral parameter $m<1/8$, applied to the operator
\begin{equation}
\edeltanot(B+\delta P_\Obl)\edeltanot
\label{65.1}
\end{equation}
viewed as an operator on the Hilbert space $\h_\Delta^0$, is
given by 
\begin{eqnarray}
\lefteqn{
F_{\Pi,m}(\edeltanot(B+\delta P_\Obl)\edeltanot)=
\edeltanot\Pi\Big( B+\delta P_\Obl}\nonumber\\
&& -(B+\delta P_\Obl) \edeltanot\Pibar
  \left(\overline{B+\delta P_\Obl}-m\right)^{-1} \Pibar \edeltanot(B+\delta
  P_\Obl)\Big) \Pi\edeltanot,\ \ \ \ \ 
\label{66}
\end{eqnarray}
where the barred operator is understood to be restricted to the subspace $\ran
\edeltanot\Pibar\subset \h_\Delta^0$. Using the definition of $A_0$, \fer{201}, and $\Pi
I_1\Pi=0$, one sees that 
\begin{equation}
\Pi B \Pi = 2\theta\lambda^2\Pi I\repsilonbar^2 I\Pi\geq 0.
\label{66'}
\end{equation}
We show that the second term on the r.h.s. of \fer{66}, which is
negative-definite, is smaller than $\Pi B\Pi$. 
By \fer{65}, the norm of the resolvent in \fer{66} is
bounded from above by $8$ (for $m< 1/8$). 
Using this fact, the estimates $\|L_0\repsilonbar\|\leq 1$,
$\|\repsilonbar^2\|\leq \epsilon^{-2}$ and $\Pibar i[L_\lambda,A_0]\Pi=\theta\lambda\Pibar
L_\lambda\repsilonbar^2I\Pi$, we find that, for any $\psi\in\h_\Delta^0$, the modulus of the expectation value 
$\av{\cdot}_\psi=\scalprod{\psi}{\cdot\, \psi}$ of the second
term in the r.h.s. of \fer{66} is bounded above by
\begin{eqnarray}
\lefteqn{
8\| \edeltanot\Pibar (\lambda
I_1 +i[L_\lambda,A_0]+\delta P_\Obl)\Pi\psi\|^2}\nonumber\\
&\leq& 16\theta^2\lambda^2 \|\repsilonbar I\Pi\psi\|^2 \nonumber\\
&&+
C\left(\delta^2\|\Pibar P_\Obl \Pi \|^2+
  C(p,\beta) \lambda^2+\frac{\theta^2\lambda^4}{\epsilon^4}\right)\|\psi\|^2.
\label{67}
\end{eqnarray}
It follows that 
\begin{eqnarray}
\lefteqn{
\av{F_{\Pi,m}(\edeltanot(B+\delta P_\Obl)\edeltanot)}_\psi}\nonumber\\
&\geq& 2\theta\lambda^2(1-8\theta)\av{\Pi I\repsilonbar^2I
  \Pi}_\psi +\delta \|P_\Obl \Pi\psi\|^2\nonumber \\
&& -C\frac{\theta\lambda^2}{\epsilon}\left(\frac{\epsilon}{\theta}C(p,\beta)+\frac{\theta\lambda^2}{
    \epsilon^3} +\frac{\epsilon}{\theta\lambda^2}\delta^2\|\Pibar
  P_\Obl \Pi\|^2\right)\|\psi\|^2. 
\label{68}
\end{eqnarray}
The expectation value on the r.h.s. of \fer{68} is estimated from below 
using
\begin{equation}
\Pi I\repsilonbar^2 I\Pi\geq \frac{1}{\epsilon}\left(\Gamma_0
  -C\epsilon^{1/4}\right),
\label{69}
\end{equation}
provided $\epsilon<\epsilon_0$, see \fer{126}, \fer{127}. Pick $\theta$ and $\epsilon$ s.t.  
\begin{equation}
\theta<1/16,\ \epsilon<\epsilon_0,
\label{parcond3}
\end{equation}
and, for $\psi\in\ran\Pi$, note the estimate
\begin{eqnarray}
\lefteqn{
\theta\lambda^2\av{I\repsilonbar^2
  I+\frac{\delta}{\theta\lambda^2} P_\Obl}_{\psi} \geq
  \frac{\theta\lambda^2}{
  \epsilon}\av{\gamma_0\Pbar_{\Omega_\beta^{at}}
  +\frac{\epsilon\delta}{\theta\lambda^2} P_\Obl
  -C\epsilon^{1/4}}_\psi}\nonumber\\
&=&\frac{\theta\lambda^2}{\epsilon}\gamma_0 \left[
  \left(1-C\epsilon^{1/4}/\gamma_0\right) \|\psi\|^2
  +\av{\frac{\epsilon\delta}{\theta\lambda^2\gamma_0}P_\Obl
    -P_\Obz}_\psi\right],
\label{71}
\end{eqnarray}
where we use that $P_{\Omega_\beta^{at}}\psi=P_\Obz\psi$ for $\psi\in\ran \Pi$.
We choose
\begin{equation}
\delta\geq \frac{\theta\lambda^2}{\epsilon}\gamma_0\geq
\frac{\theta\lambda^2}{4\epsilon} \gamma_0=: \gamma, 
\label{parcond3'}
\end{equation}
see also inequality \fer{61}, and 
\begin{equation}
C\frac{\epsilon^{1/4}}{\gamma_0}<1/4.
\label{parcond4}
\end{equation}
The r.h.s. of \fer{71} is bounded from below by
\begin{equation}
\frac{\theta\lambda^2}{\epsilon}\gamma_0\Big( 3/4- \|
P_\Obl-P_\Obz\|\Big) \|\psi\|^2\geq \frac{\theta\lambda^2}{2\epsilon}\gamma_0
\ \|\psi\|^2. 
\label{72}
\end{equation}
In the last step, we have applied Theorem \ref{betauniform}, \fer{1}, which tells us
that $\|P_\Obl-P_\Obz\|<1/4$, provided 
\begin{equation}
\mbox{$\lambda$ satisfies the condition \fer{f1} (with $\epsilon=1/4$).}
\label{cond}
\end{equation}
Combining this with \fer{68}, where we use 
\begin{equation*}
\|\Pibar P_\Obl\Pi\|^2 = 
\|\Pibar( P_\Obl-P_\Obz)\Pi\|^2\leq \|P_\Obl-P_\Obz\|^2,
\end{equation*}
 gives 
\begin{equation}
\av{F_{\Pi,m}(\edeltanot(B+\delta P_\Obl)\edeltanot)}_\psi\geq
\frac{\theta\lambda^2}{4\epsilon}\gamma_0\ \|\psi\|^2,
\label{73}
\end{equation}
provided
\begin{equation}
C\left( \frac{\epsilon}{\theta}C(p,\beta)+ \frac{\theta\lambda^2}{\epsilon^3}
+\frac{\epsilon\delta^2}{\theta\lambda^2}\right)<\gamma_0/4.
\label{parcond5}
\end{equation}
The isospectrality property of the Feshbach map tells us that
\begin{equation}
\edeltanot (B+\delta P_\Obl)\edeltanot \geq\min\left(\frac{1}{8},
  \frac{\theta\lambda^2}{4\epsilon}\gamma_0\right)\edeltanot =
  \frac{\theta\lambda^2}{4\epsilon}\gamma_0\edeltanot.
\label{74}
\end{equation}

\subsubsection{PC estimate localized w.r.t. $L_\lambda$}
\label{ss2}
Let $0\leq \chi_\Delta\leq 1$ be a smooth function with support inside the
interval $\Delta$, s.t. $\chi_\Delta(0)=1$,  and denote by $\chi_\Delta^0=\chi_\Delta(L_0)$ and $\chi_\Delta=\chi_\Delta(L_\lambda)$ the operators obtained from
the spectral theorem. We show in this subsection that any unit
vector $\psi\in\ran\Pbar_\Obl\cap {\frak B}_\nu$, s.t. $\chi_\Delta\psi=\psi$, satisfies
\begin{equation}
\av{B +\delta P_\Obl}_{\psi}= \av{B}_{\psi}\geq
\frac{\theta\lambda^2}{8\epsilon}\gamma_0,
\label{80}
\end{equation}
provided suitable bounds on the parameters $\epsilon, \lambda, \theta$ are
satisfied. We will repeatedly use the estimate 
\begin{eqnarray}
\|(1-\chi_\Delta^0)\psi\| =\|(\chi_\Delta-\chi_\Delta^0)\psi\|&\leq&
C|\lambda| \ \|I(N+1)^{-1/2}\|\ \|(N+1)^{1/2}\psi\|\nonumber\\
&\leq& C \nu |\lambda|,
\label{81}
\end{eqnarray}
where the first inequality is a consequence of the standard functional calculus. Let us decompose the expectation value
\begin{eqnarray}
\av{B}_{\psi}&=& \av{\chideltanot (B+\delta
  P_\Obl)\chideltanot}_{\psi} \label{82}\\
&& + \av{(1-\chideltanot)(B+\delta
  P_\Obl)(1-\chideltanot)}_{\psi} \label{83}\\
&&+ 2\, {\rm Re}\av{(1-\chideltanot)(B+\delta P_\Obl)\chideltanot}_{\psi}.
\label{84}
\end{eqnarray}
Because $\edeltanot\chideltanot=\chideltanot$, inequality \fer{74} implies that 
\begin{equation}
\av{\chideltanot (B+\delta P_\Obl)\chideltanot}_{\psi} \geq
\frac{\theta\lambda^2}{4\epsilon}\gamma_0 \| \chideltanot\psi\|^2
\geq 
\frac{\theta\lambda^2}{4\epsilon}\gamma_0(1-C\nu|\lambda|) \|\psi\|^2.
\label{85}
\end{equation}
Since $N+\delta P_\Obl$ is non-negative, we have that 
\begin{eqnarray}
\fer{83} &\geq& 
- \left| \av{(1-\chideltanot)(\lambda I_1 +i[L_\lambda,A_0])
    (1-\chideltanot)}_{\psi} \right| \nonumber\\
&\geq& -|\lambda|\ \|(1-\chideltanot)\psi\|\ \|I_1(N+1)^{-1/2}\|\
\|(N+1)^{1/2}\psi\| \nonumber\\
&& -\|[L_\lambda,A_0]\|\ \|(1-\chideltanot)\psi\|^2\nonumber\\
&\geq&- C\nu^2 \frac{\theta\lambda^2}{\epsilon} \left( \frac{\epsilon}{\theta}C(p,\beta) +
  |\lambda|+\frac{\lambda^2}{\epsilon}\right),
\label{86}
\end{eqnarray}
where we have used \fer{202}. \\
\indent
Our next task is to estimate \fer{84}. Since $N$ commutes (strongly) with
$\chideltanot$ and $P_\Obl\psi=0$, and using that $\psi\in\ran\Pbar_{\Obl}$, we conclude that 
\begin{eqnarray}
\lefteqn{
{\rm Re}\av{(1-\chideltanot)(B+\delta
  P_\Obl)\chideltanot}_{\psi}}\nonumber\\
&\geq& \delta \av{(1-\chideltanot)P_\Obl
  (\chideltanot-1)}_{\psi}
+{\rm Re} \av{(1-\chideltanot)(\lambda I_1
  +i[L_\lambda,A_0])\chideltanot}_{\psi}  \nonumber\\
&\geq& -\delta\|(1-\chideltanot)\psi\|^2 -C(p,\beta)\nu\lambda^2 \|\psi\|^2
-\left| \av{(1-\chideltanot)[L_\lambda,A_0]\chideltanot}_{\psi}\right|.
\label{87}
\end{eqnarray}
Taking into account that $(1-\chideltanot)\Pi=0$ and $\|(1-\chideltanot)
L_0^{-1}\|\leq C$ (the constant is of the size $|\Delta|^{-1}$), one sees
that the last term can be estimated as follows:
\begin{eqnarray}
\lefteqn{
\left|
  \av{(1-\chideltanot)[L_\lambda,A_0]\chideltanot}_{\psi}\right|}\nonumber\\
&=&\theta|\lambda|\, \left| \av{(1-\chideltanot)(\lambda I\Pi I\repsilonbar^2
 -L_\lambda\repsilonbar^2 I\Pi +\lambda \repsilonbar^2
 I\Pi I)\chideltanot}_{\psi}  \right| \nonumber\\
&&\leq C\nu\theta|\lambda|\left( \frac{\lambda^2}{\epsilon^2} +
  |\lambda|\right)\|\psi\|^2 =C\nu\frac{\theta\lambda^2}{\epsilon} \left(
  \frac{|\lambda|}{\epsilon} +\epsilon\right) \|\psi\|^2. 
\label{88}
\end{eqnarray}
Plugging \fer{88} into \fer{87} and combining this with \fer{85}, \fer{86}, we
arrive at the bound 
\begin{eqnarray}
\lefteqn{
\av{B}_{\psi}\geq}\label{89}\\
&& \frac{\theta\lambda^2}{4\epsilon}\gamma_0 \left(
  (1-C\nu|\lambda|) -\frac{C\nu}{\gamma_0}\left(\nu\frac{\epsilon}{\theta}C(p,\beta) +\nu |\lambda| +\nu\frac{\lambda^2}{\epsilon} 
  +\frac{|\lambda|}{\epsilon} +\epsilon\right)\right)\, \|\psi\|^2.
\nonumber
\end{eqnarray}
Inequality \fer{80} then follows by choosing parameters s.t.  
\begin{equation}
C\nu|\lambda|<1/4 \mbox{\ \ \ and\ \ \ } \frac{C\nu}{\gamma_0}\left(\nu\frac{\epsilon}{\theta}C(p,\beta) +\nu |\lambda| +\nu\frac{\lambda^2}{\epsilon} 
  +\frac{|\lambda|}{\epsilon} +\epsilon\right) <1/4.
\label{parcond6}
\end{equation}

\subsubsection{Choice of $\epsilon$, $\theta$ and $\delta$}
\label{ss3}
We must show that the conditions 
\begin{equation}
\mbox{\fer{parcond1}, \fer{parcond2}, \fer{parcond3},
\fer{parcond3'}, \fer{parcond4}, \fer{cond}, \fer{parcond5}, \fer{parcond6}}
\label{conditions}
\end{equation}
 can be simultaneously satisfied. 
We set 
\begin{eqnarray}
\lambda&=&\nu^{-9/2}\lambda', \label{89.9}\\
\epsilon&=&\nu^{-3}|\lambda'|^e, \mbox{\ \ \ \ some $0<e<1$,}\label{90}\\
\theta&=& |\lambda'|^t, \mbox{\ \ \ \ some $0<t<e<1$ s.t. $t>3e-2$,}\label{91}\\
\delta&=&\frac{\theta\lambda^2}{\epsilon}\gamma_0,
\label{92}
\end{eqnarray} 
and it is easily verified that there is a $\lambda_1>0$, depending on $e,t$, but not on $\nu$, $\beta\geq\beta_0$, s.t. if
\begin{equation}
0<|\lambda|<\lambda_1\min\left( C(p,\beta)^{-1}, \nu^{1/\eta-9/2}C(p,\beta)^{-1/\eta}\right),
\label{f2}
\end{equation}
where $\eta=e-t>0$, 
then conditions \fer{conditions} are met. 
The ``gap of the positive commutator'' (see \fer{80}) is of size
$\frac{\theta\lambda^2}{\epsilon}=|\lambda|^{2-\eta}\nu^{3-9\eta/2}$. The
maximal value of $\eta$ under conditions \fer{90}, \fer{91} is taken for 
$e\rightarrow 2/3$, $t\rightarrow 0$.

\section{Proof of Theorem \ref{betauniform}}
\label{betauniformsect}
The following {\it high-temperature} result is well known.  Given any
$\epsilon>0$, there is an $\eta(\epsilon)>0$ s.t. if 
\begin{equation}
\beta|\lambda|<\eta(\epsilon)
\label{47'}
\end{equation}
then inequality \fer{1} in Theorem \ref{betauniform} holds. A proof of this fact can be given by using the explicit
expression \fer{Obl} for the perturbed KMS state, and using the Dyson
series expansion to estimate $\|\Obl-\Obz\|$ (see e.g. [BFS]). Condition \fer{47'} comes from the fact that
the term of order $\lambda^n$ in the Dyson series is given by an integral over
an $n$-fold simplex of size $\beta$, and, naively, \fer{47'} is needed to ensure that $\|\Obl-\Obz\|$ is small. We shall improve our estimates on $\|\Obl-\Obz\|$ by taking advantage of the decay in (imaginary) time of the field propagators. \\

To start our analysis, we use the fact that the trace-norm 
majorizes the operator-norm to write 
\begin{eqnarray}
\|P_\Obl-P_\Obz\|^2&\leq& \|P_\Obl-P_\Obz\|^2_2=2\left( 1-\scalprod{\Obl}{P_\Obz
    \Obl}\right)\nonumber\\
&\leq& 2\scalprod{\Obl}{ \Pbar_{\Omega_\beta^{at}}
  \Obl} +2\scalprod{\Obl}{ \Pbar_\Omega \Obl},
\label{47}
\end{eqnarray}
where we use $\bbbone- P_\Obz\leq \Pbar_{\Omega_\beta^{at}}
+\Pbar_\Omega$. Here, $\Omega_\beta^{at}$ is the atomic Gibbs state at inverse
temperature $\beta$ given in \fer{111}, and $\Omega$ is the vacuum vector in $\cal F$, see \fer{109}. We know that
\begin{equation}
\scalprod{\Obl}{ \Pbar_\Omega \Obl}\leq \|N^{1/2}\Obl\|^2\leq
c(p,\beta)^2 |\lambda|^2,
\label{48}
\end{equation}
where $c(p,\beta)$ satisfies \fer{cpb1}, see Theorem \ref{vthm}. There is a $\beta_1(\epsilon)\geq \beta_0$ s.t. if $\beta>\beta_1(\epsilon)$ then 
\begin{equation}
\|P_{\Omega_\beta^{at}}-P_{\varphi_0\otimes\varphi_0}\| < \epsilon/2,
\label{49}
\end{equation}
where $\varphi_0$ is the groundstate
eigenvector of $H_{at}$ and $P_{\varphi_0\otimes\varphi_0}\in{\cal B}(\h_{at}\otimes \h_{at})$ is the projection onto the span of $\varphi_0\otimes\varphi_0$. It follows from \fer{47} that
\begin{equation}
\|P_\Obl-P_\Obz\|^2 \leq 2\scalprod{\Obl}{\Pbar_{\!\!{\varphi_0\otimes \varphi_0}}\Obl}+\epsilon +2c(p,\beta)^2 |\lambda|^2,
\label{50'}
\end{equation}
for $\beta>\beta_1(\epsilon)$. Let 
\begin{equation}
Q=\Pbar_{\varphi_0}\in{\cal B}(\h_{at})
\label{49'}
\end{equation}
be the projection onto the
orthogonal complement 
of the groundstate subspace of the atomic Hamiltonian $H_{at}$ so that 
\begin{equation}
\Pbar_{\varphi_0\otimes\varphi_0}\leq Q\otimes\bbbone_{at} + \bbbone_{at}\otimes Q. 
\label{51}
\end{equation}
Noticing that 
$\scalprod{\Obl}{Q\otimes\bbbone_{at}\
  \Obl}=\scalprod{\Obl}{\bbbone_{at}\otimes Q\ \Obl}=\obl(Q)$ we see from \fer{50'} that 
\begin{equation}
\|P_\Obl-P_\Obz\|^2 \leq 4\obl(Q)+\epsilon +2c(p,\beta)^2 |\lambda|^2,
\label{50}
\end{equation}  
provided $\beta>\beta_1(\epsilon)$. 
\begin{proposition}
\label{prop1}
For any $\epsilon>0$ there exist $\beta_2(\epsilon)>0$ and
$\lambda_1(\epsilon)>0$ such that if $\beta > \beta_2(\epsilon)$ and
$|\lambda|<\lambda_1(\epsilon)$ then 
\begin{equation}
\obl(Q)<\epsilon.
\label{2}
\end{equation}
\end{proposition}
The proof is presented below. For now, we use \fer{2} to prove Theorem
\ref{betauniform}. We set
\begin{eqnarray}
\beta_3(\epsilon)&:=&\max(\beta_1(\epsilon), 
\beta_2(\epsilon)),\nonumber\\ 
\lambda'_0(\epsilon)&:=&\min\left(\lambda_1(\epsilon),
c(p,\beta)^{-1}\sqrt{\epsilon/2},\eta(\epsilon)/\beta_3(\epsilon) \right),
\label{52}
\end{eqnarray}
where $\eta(\epsilon)$ is the constant appearing in \fer{47'}. In the case $p>-1/2$ the constant $c(p,\beta)$ has an upper bound which is uniform in $\beta\geq \beta_0$, see \fer{cpb1}, and we take $\lambda_0(\epsilon)$ to be the r.h.s. of \fer{52} with $c(p,\beta)$ replaced by this upper bound. For $p=-1/2$ we can find a $\lambda_0(\epsilon)$, indpendent of $\beta>0$, satisfying $(1+\log(1+\beta))^{-1}\lambda_0(\epsilon)\leq \lambda_0'(\epsilon)$, see \fer{cpb1}. \\
\indent
We always assume \fer{f1}. Inequalities \fer{50} and
\fer{2} yield
\begin{equation}
\|P_\Obl-P_\Obz\|^2 \leq 6\epsilon,
\label{53}
\end{equation}
for $\beta>\beta_3(\epsilon)$.  
If $\beta\leq\beta_3(\epsilon)$ then $\beta|\lambda|< \eta(\epsilon)$, and
\fer{1} follows from the high-temperature result mentioned above. This completes
the proof of the theorem, given Proposition \ref{prop1}.\\
\indent
{\it Proof of Proposition \ref{prop1}.\ } It is convenient to work with a finite volume approximation 
\begin{equation}
\oblL(\cdot)=\frac{\tr \left(e^{-\beta\HlL}\,\cdot\,\right)}{\tr e^{-\beta \HlL}}
\label{3}
\end{equation}
of the KMS state $\obl$, where $\Lambda= [-L/2,L/2]^3\subset\r^3$. (We introduce a finite box $\Lambda$ just in order to be able to make use of some familiar inequalities for traces. The inequalities needed in our proof also hold in the thermodynamic limit, $\Lambda\nearrow \r^3$; but some readers may be less familiar with them.) In \fer{3}, the trace
is taken over the Hilbert space $\h_{at}\otimes {\cal F}(L^2(\Lambda,d^3x))$. 
For $n=(n_1,n_2,n_3)\in{\mathbb
  Z}^3$, let  
\begin{equation}
e^\Lambda_n(x)=L^{-3/2}e^{2\pi i n x/L}, \
E_n^\Lambda=\frac{2\pi}{L}|n|=\frac{2\pi}{L}(n_1^2+n_2^2+n_3^2)^{1/2}
\label{6}
\end{equation}
denote the eigenvectors and eigenvalues of the operator $\sqrt{-\Delta}$ on $L^2(\Lambda,d^3x)$ with periodic boundary conditions at $\partial \Lambda$. We identify the basis $\{e^\Lambda_n\}$ of $L^2(\Lambda^3,d^3x)$ with the canonical basis of $l^2({\mathbb Z}^3)$, and define the finite-volume Hamiltonian by
\begin{eqnarray}
\HlL&=& H_{at}+\HfL+\lambda v^\Lambda,\label{4}\\
v^\Lambda&=&\sum_\alpha G_\alpha\otimes\varphi(g_\alpha^\Lambda),\label{5}
\end{eqnarray}
where $g_\alpha^\Lambda\in l^2({\mathbb Z}^3)$ is given by 
\begin{equation}
g_\alpha^\Lambda(n)=\left(\frac{2\pi}{L}\right)^{3/2} 
\left\{
\begin{array}{ll}
g_\alpha\left(\frac{2\pi n}{L}\right), & n\neq 0,\\
1,& n=0.
\end{array}
\right.
\label{6.1}
\end{equation}
and the operator
\begin{equation}
\HfL=\d\Gamma(\hfL),
\label{7}
\end{equation}
acting on ${\cal F}(l^2({\mathbb Z}^3))$, is the second quantization of the one-particle Hamiltonian 
\begin{equation}
\hfL e_n^\Lambda=\left\{
\begin{array}{cl}
E_n^\Lambda e_n^\Lambda, & \mbox{if $n\neq(0,0,0)$},\\
e_n^\Lambda, & \mbox{if $n=(0,0,0)$}.
\end{array}
\right.
\label{8}
\end{equation}
On the complement of the zero-mode subspace $\hfL$ equals $\sqrt{-\Delta}$
 with periodic boundary conditions. Changing the action of $\hfL$ on 
 finitely many modes (always under the condition that $e^{-\beta \HfL}$ is
 trace-class) does not affect the thermodynamic limit. Similarly, we may modify the definition of $g_\alpha^\Lambda$ on finitely many modes without altering the thermodynamic limit. The existence of the thermodynamic limit, 
\begin{equation}
\lim_{L\rightarrow\infty} \oblL(A)=\obl(A),
\label{9}
\end{equation}
can be proven by expanding $e^{-\beta \HlL}$ into a Dyson
(perturbation) series and using that 
\begin{equation}
\obzL(A)=\frac{\tr\left( e^{-\beta\HzL}A \right)}{\tr e^{-\beta
    \HzL}}
\label{9.1}
\end{equation}
has the expected thermodynamic limit for quasi-local observables $A$. \\
\indent
Our goal is to show that 
$\oblL(Q)<\epsilon$, for $Q$ given in \fer{49'}, provided $\beta$ and $\lambda$ satisfy the conditions given in Proposition \ref{prop1}, {\it uniformly} in the size of $\Lambda$. In what follows, we
will use the H\"older and Peierls-Bogoliubov inequalities (see
e.g. [S]). The H\"older inequality (for traces) reads 
\begin{equation}
\|A_1\ldots A_n\|_1\leq\prod_{j=1}^n \|A_j\|_{p_j},
\label{holder}
\end{equation}
where $1\leq p_j\leq\infty$, $\sum_j p_j^{-1}=1$, and the norms are  
\begin{equation}
\|A\|_p=\left(\tr |A|^p\right)^{1/p}, \mbox{for $p<\infty$, and }
\|A\|_\infty=\|A\| \mbox{\ (operator norm)}.
\label{10}
\end{equation}
The Peierls-Bogoliubov inequality says that
\begin{equation}
\frac{\tr\left(e^{A+B}\right)}{\tr e^B}\geq \exp\left[ \tr \left(Ae^B\right)/\tr e^B\right],  
\label{peierls}
\end{equation}
which implies that 
\begin{equation}
\frac{\tr e^{-\beta\HzL}}{\tr e^{-\beta\HlL}}\leq
e^{\beta|\lambda \obzL(v^\Lambda)|}=1,
\label{applipeierls}
\end{equation}
since, by \fer{5}, $\obzL(v^\Lambda)=0$.\\ 
\indent
Using the H\"older inequality one sees that, for any
$0<\tau\leq \beta/2$,
\begin{eqnarray}
\lefteqn{
\oblL(Q)=\frac{\tr\left(e^{-(\beta-2\tau)\HlL}e^{-\tau\HlL}Qe^{-\tau
      \HlL}\right) }{\tr e^{-\beta\HlL}}}\nonumber\\
&\leq& \left[\frac{\tr \left\{\left(e^{-\tau\HlL}Qe^{-\tau\HlL
      }\right)^{\frac{\beta}{2\tau}}\right\}}{\tr e^{-\beta\HlL}}\right]
      ^{\frac{2\tau}{\beta}}
= \left[\frac{\tr \left\{\left(Qe^{-\frac{\beta}{2M}\HlL}Q
      \right)^{2M}\right\}}{\tr e^{-\beta\HlL}}\right] 
      ^{\frac{1}{2M}}\!\!\!,\ \ \ \ \ \ \ \ 
\label{11}
\end{eqnarray}
where we are choosing $\tau$ s.t.  
\begin{equation}
\frac{\beta}{2\tau}=2M,\ \ \mbox{for some $M\in{\mathbb N}$}.
\label{12}
\end{equation}
Setting 
\begin{equation}
v^\Lambda(t)=e^{-t\HzL}v^\Lambda e^{t\HzL}
\label{12'}
\end{equation}
and using the Dyson series expansion we obtain
\begin{equation}
Q e^{-\frac{\beta}{2M}\HlL}Q = A+B,
\label{13}
\end{equation}
where the selfadjoint operators $A$ and $B$ are given by 
\begin{eqnarray}
A&=& Qe^{-\frac{\beta}{2M}\HzL}Q\label{14}\\
B&=& \sum_{n\geq 1}
(-\lambda)^n\int_{0\leq t_n\leq \ldots\leq t_1\leq
  \frac{\beta}{2M}} Qv^\Lambda(t_n)\cdots
v^\Lambda(t_1)e^{-\frac{\beta}{2M}\HzL} Q\  dt_1\cdots dt_n.\ \ \ \ \ \ \ \ 
\label{15}
\end{eqnarray}
We plug \fer{13} into \fer{11}, expand $(A+B)^{2M}$ and use the H\"older
inequality to arrive at the bound
\begin{equation}
\oblL(Q)\leq \left[ \frac{\tr \left(|A|^{2M}\right)}{\tr
    e^{-\beta\HlL}}\right]^{\frac{1}{2M}} + 
\left[ \frac{\tr \left(|B|^{2M}\right)}{\tr
    e^{-\beta\HlL}}\right]^{\frac{1}{2M}}.
\label{16}
\end{equation}
The first term on the right hand of \fer{16} is easy to estimate. Let
$\Delta=E_1-E_0>0$ denote the spectral gap of the atomic Hamiltonian $H_{at}$.
Then
\begin{eqnarray}
\frac{\tr\left( |A|^{2M}\right)}{\tr e^{-\beta\HzL}}&=& 
\frac{\tr_{\h_{at}}\left( Qe^{-\beta H_{at}}\right)}{\tr_{\h_{at}} e^{-\beta H_{at}}}
=\frac{\sum_{j=1}^{d-1} e^{-\beta(E_j-E_0)}}{1+\sum_{j=1}^{d-1} e^{-\beta
    (E_j-E_0)}}\nonumber\\
&\leq& \sum_{j=1}^{d-1} e^{-\beta(E_j-E_0)}\leq
2\int_{E_1-E_0}^\infty e^{-\beta x}dx=2\frac{e^{-\beta\Delta}}{\beta}.
\label{17}
\end{eqnarray}
Taking into account \fer{applipeierls} and \fer{12}, we obtain, for $\beta\geq 1$, 
\begin{equation}
\left[ \frac{\tr \left(|A|^{2M}\right)}{\tr
    e^{-\beta\HlL}}\right]^{\frac{1}{2M}}\leq
    2e^{-2\tau\Delta}.
\label{18}
\end{equation}
In order to
make the r.h.s. small, we take $\tau$ large  as compared to $\Delta^{-1}$ (hence $\beta\geq 2\tau$ 
must be large enough). \\
\indent
Next, we consider the second term on the r.h.s. of \fer{16}. From
\fer{applipeierls} one sees that
\begin{equation}
\frac{\tr\left( |B|^{2M}\right)}{\tr e^{-\beta \HlL}}\leq \obzL\left(
  e^{\beta\HoL} |B|^{2M}\right)=\obzL\left( e^{\beta\HoL} B^{2M}\right).
\label{19}
\end{equation}
We expand
\begin{equation}
e^{\beta\HoL} B^{2M} =\sum_{k_1,\ldots, k_{2M}\geq 1} T(k_1,\ldots,k_{2M}),
\label{20}
\end{equation}
where 
\begin{eqnarray}
\lefteqn{
T(k_1,\ldots,k_{2M}) =(-\lambda)^{k_1+\cdots +k_{2M}}
\int_0^{\frac{\beta}{2M}} dt^{(1)}_1\cdots \int_0^{t^{(1)}_{k_1-1}}
dt^{(1)}_{k_1}} \nonumber\\
&&
\times
\int_{\frac{\beta}{2M}}^{2\frac{\beta}{2M}}dt^{(2)}_1\cdots
\int_{\frac{\beta}{2M}}^{t^{(2)}_{k_2-1}} dt_{k_2}^{(2)} 
\ \cdots\ \int_{(2M-1)\frac{\beta}{2M}}^\beta dt_1^{(2M)}\cdots
\int_{(2M-1)\frac{\beta}{2M}}^{t_{k_{2M}-1}^{(2M)}} dt_{k_{2M}}^{(2M)}
\nonumber\\
&&
\times
e^{\beta\HoL} Q v^\Lambda(t_{k_1}^{(1)})\cdots
v^\Lambda(t_1^{(1)})Q \ 
Qv^\Lambda(t^{(2)}_{k_2})\cdots v^\Lambda(t_1^{(2)}) Q\times \cdots \nonumber\\
&&
\cdots \times Qv^\Lambda(t^{(2M)}_{k_{2M}})\cdots v^\Lambda(t_1^{(2M)}) Q
e^{-\beta\HoL}.
\label{21}
\end{eqnarray}
Note that the time variables in the integrand are ordered,
\begin{equation}
0\leq t_{k_1}^{(1)}\leq\cdots\leq t_1^{(1)}\leq t_{k_2}^{(2)}\leq\cdots
\leq t_1^{(2M)}\leq \beta.
\label{22'}
\end{equation}
Our goal is to obtain an upper bound on $|\obzL(T(k_1,\ldots,k_{2M}))|$, sharp enough to show that 
\begin{equation}
\sum_{k_1,\ldots,k_{2M}\geq 1} \left|\obzL(T(k_1,\ldots,k_{2M}))\right| 
\label{22}
\end{equation}
converges, and to estimate the value of the series. Note that the factors $e^{\beta \HoL}$ and $e^{-\beta\HoL}$ 
in the integrand in \fer{21}  drop when we apply $\obzL$
(cyclicity of the trace), and the expectation value of the integrand in the state
$\obzL=\omega_\beta^{at}\otimes \omega_\beta^{f,\Lambda}$ (see \fer{9.1}) splits into a sum over products
\begin{eqnarray}
\lefteqn{
\sum_{\alpha_1^{(1)},\ldots,\alpha_{k_1}^{(1)}}\cdots
\sum_{\alpha_1^{(2M)},\ldots,\alpha_{k_{2M}}^{(2M)}} \omega_\beta^{at}\left(
  QG_{\alpha_{k_1}^{(1)}}(t^{(1)}_{k_1}) \cdots 
G_{\alpha_1^{(2M)}}(t^{(2M)}_1)Q\right)}\nonumber\\
&&\ \ \ \ \ \ \ \ \ \times \omega_\beta^{f,\Lambda}\left(
  \varphi_{\alpha_{k_1}^{(1)}}^\Lambda(t_{k_1}^{(1)})\cdots 
\varphi_{\alpha_1^{(2M)}}^\Lambda(t_1^{(2M)})\right),\ \ \ \ \ \ \ \ \ 
\label{23}
\end{eqnarray}
where $\omega_\beta^{at}$ and $\omega^{f,\Lambda}_\beta$ are the atomic and
field KMS states at inverse temperature $\beta$, and 
\begin{eqnarray}
G_\alpha(t)&=&e^{-tH_{at}}G_\alpha e^{tH_{at}}\label{24}\\
\varphi_\alpha^\Lambda(t)&=&e^{-t\HfL}\varphi(g^\Lambda_\alpha)e^{t\HfL} =
a^*\left(e^{-t\hfL}g_\alpha^\Lambda\right) +a\left(e^{t\hfL}
  g_\alpha^\Lambda\right).
\label{25}
\end{eqnarray}
Using the H\"older inequality \fer{holder} it is not difficult to see that 
\begin{equation}
\left| \omega^{at}_\beta\left(
  QG_{\alpha_{k_1}^{(1)}}(t^{(1)}_{k_1}) \cdots 
G_{\alpha_1^{(2M)}}(t^{(2M)}_1)Q\right)\right|\leq \prod_{j=1}^{2M}
  \|G_{\alpha_1^{(j)}}\| \cdots \|G_{\alpha_{k_j}^{(j)}}\|.
\label{26}
\end{equation}
Since $\omega^{f,\Lambda}_\beta$ is a quasi-free state 
we can estimate the second factor in \fer{23} with the help of
Wick's theorem: 
\begin{equation}
\omega_\beta^{f,\Lambda} \left( \varphi_{\alpha_1}^\Lambda (t_1) \cdots \varphi_{\alpha_{2N}}^\Lambda (t_{2N})\right) 
=\sum_{\cal P} \prod_{(l,r)\in{\cal P}} \omega_\beta^{f,\Lambda} \left( \varphi_{\alpha_l}^\Lambda (t_l) \varphi_{\alpha_{r}}^\Lambda (t_{r})\right),
\label{26.1}
\end{equation}
where the sum extends over all {\it contraction schemes}, i.e., decompositions of $\{1,\ldots,2N\}$ into $N$ disjoint, ordered pairs $(l,r)$, $l<r$. Applying  \fer{26.1} to 
\begin{equation}
\omega_\beta^{f,\Lambda} \left(
  \varphi_{\alpha_{k_1}^{(1)}}^\Lambda(t_{k_1}^{(1)})\cdots 
\varphi_{\alpha_1^{(2M)}}^\Lambda(t_1^{(2M)})\right)
\label{27}
\end{equation}
we find that all resulting terms can be organized in {\it graphs} $\cal G$, constructed in the following way. Partition the circle of circumference
$\beta$ into $2M$ segments (parametrized by the arc length)
$\Delta_j=[(j-1)\frac{\beta}{2M}, j\frac{\beta}{2M}]$, $j=1,\ldots,2M$. Put
$k_j$ ``dots'' into the interval $\Delta_j$, each dot representing a time
variable $t^{(j)}_\cdot\in \Delta_j$ (increasing times are ordered according
to increasing arc length). Pick any dot in any interval and pair it
with an arbitrary different dot in any interval. Then pick
any unpaired dot (i.e., one not yet paired up) and pair it with any other unpaired dot. Continue this procedure until all dots in all intervals are paired;
 (notice that the total number of dots on the circle is even, as
follows from the gauge-invariance of $\omega^{f,\Lambda}_\beta$). The graph
$\cal G$ associated to such a pairing consists of all pairs -- including
multiplicity -- of 
intervals $(\Delta,\Delta')$ with the property that some dot in $\Delta$ is
paired with some dot in $\Delta'$. ``Including multiplicity'' means that if, say, three dots of $\Delta$ are paired with three dots in
$\Delta'$, we understand that $\cal G$ contains the pair $(\Delta, \Delta')$
three times. The class of all pairings $\cal P$ leading to a given graph $\cal G$ is denoted by $C_{\cal G}$. Let 
\begin{equation}
A_{\cal P}=\prod_{(l,r)\in{\cal P}} \omega_\beta^{f,\Lambda} \left( \varphi_{\alpha_l}^\Lambda (t_l) \varphi_{\alpha_{r}}^\Lambda (t_{r})\right)
\label{26.2}
\end{equation}
denote the contribution to \fer{26.1} corresponding to the pairing $\cal P$. The numerical value, $|{\cal G}|$, corresponding to a graph $\cal G$ is defined by
\begin{equation}
|{\cal G}|=\left| \sum_{{\cal P}\in C_{\cal G}} A_{\cal P}\right|,
\label{26.3}
\end{equation}
and it follows from \fer{26.1}, \fer{26.2} and \fer{26.3} that 
\begin{equation}
\left|\omega^\Lambda_{\beta,f}\left(\varphi_{\alpha_{k_1}^{(1)}}^\Lambda(t_{k_1}^{(1)})\cdots 
\varphi_{\alpha_1^{(2M)}}^\Lambda(t_1^{(2M)}) \right)\right| \leq \sum_{\cal
G}|{\cal G}|.
\label{28}
\end{equation}
In order to give an upper bound on the r.h.s. of \fer{28}, we must estimate the imaginary-time propagators (two-point functions)
\begin{eqnarray}
\lefteqn{
\omega_\beta^{f,\Lambda}\left( e^{-t_l \HfL} \varphi(g_{\alpha_l}^\Lambda) e^{t_l \HfL}
e^{-t_r\HfL} \varphi(g_{\alpha_r}^\Lambda)e^{t_r \HfL}\right)}\nonumber\\
&=&\scalprod{g_{\alpha_r}^\Lambda}{e^{-(\beta +t_l-t_r)\hfL}\frac{e^{\beta \hfL}}{e^{\beta\hfL}-1} g_{\alpha_l}^\Lambda}
+\scalprod{g_{\alpha_l}^\Lambda}{e^{-(t_r-t_l)\hfL}\frac{e^{\beta \hfL}}{e^{\beta\hfL}-1} g_{\alpha_r}^\Lambda}
\label{28.1}\ \ \ \ \ 
\end{eqnarray}
where the $g_{\alpha_{l,r}}^\Lambda\in l^2({\mathbb Z}^3)$ are given in \fer{6.1}, and where $t_l\in\Delta_l$, $t_r\in \Delta_r$ s.t. $0\leq t_l\leq t_r\leq \beta$. 
The r.h.s. of \fer{28.1} equals
\begin{eqnarray}
\lefteqn{
\left(\frac{2\pi}{L}\right)^3\sum_{n\neq (0,0,0)}
  \Big[ \overline{g_{\alpha_r}(2\pi n/L)} g_{\alpha_l}(2\pi n/L)
  e^{-(\beta+t_l-t_r)E_n^\Lambda}}\nonumber\\
&&+\overline{g_{\alpha_l}(2\pi n/L)} g_{\alpha_r}(2\pi n/L) e^{-(t_r-t_l)E_n^\Lambda} \Big] \times \frac{e^{\beta E_n^\Lambda}}{e^{\beta E_n^\Lambda}-1}  \nonumber\\
&+& \left(\frac{2\pi}{L}\right)^3\left[ e^{-(\beta+t_l-t_r)} +e^{-(t_r-t_l)}\right]\frac{e^\beta}{e^\beta-1}. \ \ \ \ \ \ \ \ \ \ \ \ 
\label{32}
\end{eqnarray}
In the limit $L\rightarrow\infty$, the Riemann sum in \fer{32} converges to
\begin{equation}
\int_{\r^3} d^3k \ \left[ \overline{g_{\alpha_r}(k)} g_{\alpha_l}(k) 
  e^{-(\beta+t_l-t_r)|k|} + \overline{g_{\alpha_l}(k)} g_{\alpha_r}(k) e^{-(t_r-t_l)|k|}\right]
  \frac{e^{\beta|k|}}{e^{\beta|k|}-1},
\label{33}
\end{equation}
since the form factors $g_{\alpha_{l,r}}$ satisfy conditions (A1). The term in \fer{32} coming from $n=(0,0,0)$
disappears in the limit $L\rightarrow\infty$. (This shows why a redefinition of $\hfL$ on the zero mode
 does not affect the thermodynamic limit.)\\
\indent
It is not hard to see that, for arbitrary $\Delta_l, \Delta_r$ and
$t_l\in\Delta_l$, $t_r\in\Delta_r$, 
\begin{equation}
|t_l-t_r|\geq d_-(\Delta_l,\Delta_r):=\frac{\beta}{2M}\left\{
\begin{array}{ll}
0, & \mbox{if $l=r$}\\
|l-r|-1, & \mbox{if $l\neq r$}
\end{array}
\right.
\label{32'}
\end{equation}
and 
\begin{equation}
\beta -|t_l-t_r|\geq
d_+(\Delta_l,\Delta_r):=\beta -\frac{\beta}{2M}\left(|l-r|+1\right).
\label{33'}
\end{equation}
Defining
\begin{equation}
d(\Delta, \Delta'):=\min(d_-(\Delta, \Delta'), d_+(\Delta, \Delta')),
\label{34}
\end{equation}
we obtain from \fer{28.1} and \fer{33}, and for $L$ large enough,
\begin{eqnarray}
\lefteqn{
\left| \omega_\beta^{f,\Lambda}\left( e^{-t_l \HfL} \varphi(g^\Lambda_{\alpha_l}) e^{t_l \HfL}
e^{-t_r\HfL} \varphi(g_{\alpha_r}^\Lambda)e^{t_r \HfL}\right)\right|}\nonumber\\
&&\leq 2\scalprod{g_{\alpha_l}^\Lambda}{\frac{e^{-d(\Delta_l,
    \Delta_r)|k|}}{1-e^{-\beta|k|}} g_{\alpha_l}^\Lambda}^{1/2} 
\scalprod{g_{\alpha_r}^\Lambda}{\frac{e^{-d(\Delta_l,
    \Delta_r)|k|}}{1-e^{-\beta|k|}}g_{\alpha_r}^\Lambda}^{1/2}.
\label{35}
\end{eqnarray}
Given any two intervals $\Delta, \Delta'$, set 
\begin{equation}
C(\Delta,\Delta'):=4\max_{\alpha}\scalprod{
  g_\alpha}{\frac{e^{-d(\Delta,  \Delta')|k|}}{1-e^{-\beta|k|}}
  g_\alpha}.
\label{36}
\end{equation}
If $L \geq C$, for some constant $C$, then \fer{36}
is a volume-independent upper bound on the (finite-volume) two-point functions
arising from contractions in the graph expansion (Wick theorem). We are now
ready to give an upper bound on the r.h.s. of \fer{28}; (see also [F] for similar considerations). \\
\indent
It is useful to start the procedure of pairing dots in the interval with the
highest order $k$. Let $\pi$ be a permutation of $2M$ objects, s.t.
\begin{equation}
k_{\pi(1)}\geq k_{\pi(2)}\geq\cdots\geq k_{\pi(2M)}.
\label{29'}
\end{equation}
There are $k_{l_1^{(\pi(1))}}$ possibilities of pairing the dot
  $t^{(\pi(1))}_1$ with  
  some dot in an interval $\Delta_{l_1^{(\pi(1))}}$. We associate to each such pairing the value 
\begin{equation}
 k_{l_1^{(\pi(1))}} \ C(\Delta_{\pi(1)},\Delta_{l_1^{(\pi(1))}})\leq
\sqrt{k_{\pi(1)}} \ \sqrt{k_{l_1^{(\pi(1))}}}\ 
C(\Delta_{\pi(1)},\Delta_{l_1^{(\pi(1))}}), 
\label{30'}
\end{equation}
where we use \fer{29'}. Next, we pair the dot labelled by
$t_2^{(\pi(1))}$ (if it is still unpaired, otherwise we move to the next unpaired dot) with a dot in $\Delta_{l_2^{(\pi(1))}}$ and associate to this pairing the
value 
\begin{equation}
\sqrt{k_{\pi(1)}} \ \sqrt{k_{l_2^{(\pi(1))}}}\ 
C(\Delta_{\pi(1)},\Delta_{l_2^{(\pi(1))}}).
\label{37}
\end{equation}
We continue this procedure until all dots are paired. This yields the estimate
\begin{equation}
\sum_{\cal G}|{\cal G}|\leq\prod_{j=1}^{2M} (k_j)^{k_j/2}\sum_{\cal G}
 \prod_{(\Delta,\Delta')\in{\cal G}} 
 C(\Delta,\Delta').
\label{38}
\end{equation}
Next, we establish an upper bound on the sum on the r.h.s. Using that
\begin{equation}
|g_\alpha(k) |\leq C |k|^p, 
\label{39}
\end{equation}
for some constant $C$, and for all $\alpha$, 
provided $|k|$ is small enough, with $p>-1$, it is easy to see that 
\begin{equation}
C(\Delta,\Delta')\leq C\left\{
\begin{array}{ll}
d(\Delta,\Delta')^{-3-2p},& d(\Delta,\Delta')\neq 0\\
1/\beta +1,& d(\Delta,\Delta')=0
\end{array}
\right.
\label{40}
\end{equation}
Furthermore, using definition \fer{34} and inequality \fer{40}, we see that, for any $\Delta$, 
\begin{equation}
\sum_{\Delta'} C(\Delta,\Delta')\leq\Gamma:= 
C\left(1+\frac{1}{\beta}+\frac{1}{p+1}\left(\frac{\beta}{2M}\right)^{-2-2p}
  \right)<\infty,
\label{41}
\end{equation}
provided $p>-1$. 
Consequently, we find that 
\begin{equation}
\sum_{\cal G} \prod_{(\Delta,\Delta')\in{\cal G}}  C(\Delta,\Delta')\leq
\Gamma^{k_1+\cdots +k_{2M}}.
\label{42}
\end{equation}
Carrying out the integral over the simplex in \fer{21}, and using \fer{23},
\fer{26}, \fer{28}, \fer{38}, \fer{42}, we obtain the bound 
\begin{equation}
\left|\obzL(T(k_1,\ldots,k_{2M}))\right|\leq \left(C'|\lambda|\Gamma
  \frac{\beta}{2M}\right)^{k_1+ \cdots + k_{2M}}\  \prod_{j=1}^{2M}
  \frac{(k_j)^{k_j/2}}{k_j!}, 
\label{43}
\end{equation} 
where $C'=\sum_\alpha \|G_\alpha\|$, and where the factor $(\frac{\beta}{2M})^{k_j}\frac{1}{k_j!}$ is the volume of the simplex $\{t\leq t_{k_j}\leq \cdots\leq t_1\leq t+\frac{\beta}{2M}\}$. Thus, the series
\fer{22} converges for all values of $\lambda$ and $\beta>0$, and 
\begin{equation}
\left[ \obzL\left(e^{\beta\HoL} B^{2M}\right)\right]^{\frac{1}{2M}} \leq 
 C'|\lambda|\Gamma\frac{\beta}{2M} \sum_{k\geq 0} \left(
 C'|\lambda|\Gamma\frac{\beta}{2M}\right)^k \frac{(k+1)^{\frac{k+1}{2}}}{
 (k+1)!}. 
\label{44}
\end{equation}
Combining \fer{16}, \fer{18}, \fer{19}, \fer{44}, and using \fer{12}, we see
that if 
$L$ is large enough (independly of $\lambda$ or
$\beta$) then 
\begin{equation}
\oblL(Q)\leq 2e^{-2\tau\Delta} + C'|\lambda|\Gamma\tau \sum_{k\geq 0}
\left( C'|\lambda|\Gamma\tau\right)^k \frac{(k+1)^{\frac{k+1}{2}}}{(k+1)!}.
\label{45}
\end{equation}
The final step in the proof of Proposition \ref{prop1} consists in showing
that the 
r.h.s. (which is independent of $\Lambda$) can be made arbitrarily small, provided $\beta$ is large enough and $\lambda$ is small enough.
Pick $\beta_2(\epsilon)>1$ so large that
$e^{-\beta_2(\epsilon)\Delta}<\epsilon/2$. For $\beta\geq
\beta_2(\epsilon)$ we choose $\tau=\beta_2(\epsilon)/2\leq\beta/2$. From the
definition of $\Gamma$, \fer{41}, and the relation $\frac{\beta}{2M}=2\tau$,
see \fer{12}, we see that 
$ \Gamma\tau\leq C(\epsilon)$, uniformly in $\beta\geq\beta_2(\epsilon)$. 
It follows that there is a $\lambda_1(\epsilon)>0$ s.t. if
$|\lambda|<\lambda_1(\epsilon)$ then the second term on the r.h.s. of
\fer{45} is smaller than $\epsilon/2$. 
This completes the proof of Proposition \ref{prop1}.\hfill $\blacksquare$

\end{document}